\documentclass[12pt]{article}
\usepackage{graphics}

\def\hybrid{\topmargin -20pt    \oddsidemargin 0pt
        \headheight 0pt \headsep 0pt
        \textwidth 6.25in       
        \textheight 9.5in       
        \marginparwidth .875in
        \parskip 5pt plus 1pt   \jot = 1.5ex}

\hybrid

\def\baselinestretch{1.2}

\catcode`\@=11

\def\marginnote#1{}
\newcount\hour
\newcount\minute
\newtoks\amorpm
\hour=\time\divide\hour by60
\minute=\time{\multiply\hour by60 \global\advance\minute by-\hour}
\edef\standardtime{{\ifnum\hour<12 \global\amorpm={am}%
        \else\global\amorpm={pm}\advance\hour by-12 \fi
        \ifnum\hour=0 \hour=12 \fi
        \number\hour:\ifnum\minute<10 0\fi\number\minute\the\amorpm}}
\edef\militarytime{\number\hour:\ifnum\minute<10 0\fi\number\minute}

\def\draftlabel#1{{\@bsphack\if@filesw {\let\thepage\relax
   \xdef\@gtempa{\write\@auxout{\string
      \newlabel{#1}{{\@currentlabel}{\thepage}}}}}\@gtempa
   \if@nobreak \ifvmode\nobreak\fi\fi\fi\@esphack}
        \gdef\@eqnlabel{#1}}
\def\@eqnlabel{}
\def\@vacuum{}
\def\draftmarginnote#1{\marginpar{\raggedright\scriptsize\tt#1}}

\def\draft{\oddsidemargin -.5truein
        \def\@oddfoot{\sl preliminary draft \hfil
        \rm\thepage\hfil\sl\today\quad\militarytime}
        \let\@evenfoot\@oddfoot \overfullrule 3pt
        \let\label=\draftlabel
        \let\marginnote=\draftmarginnote
   \def\@eqnnum{(\theequation)\rlap{\kern\marginparsep\tt\@eqnlabel}%
\global\let\@eqnlabel\@vacuum}  }

\def\preprint{\twocolumn\sloppy\flushbottom\parindent 2em
        \leftmargini 2em\leftmarginv .5em\leftmarginvi .5em
        \oddsidemargin -.5in    \evensidemargin -.5in
        \columnsep .4in \footheight 0pt
        \textwidth 10.in        \topmargin  -.4in
        \headheight 12pt \topskip .4in
        \textheight 6.9in \footskip 0pt
        \def\@oddhead{\thepage\hfil\addtocounter{page}{1}\thepage}
        \let\@evenhead\@oddhead \def\@oddfoot{} \def\@evenfoot{} }

\def\numberbysection{\@addtoreset{equation}{section}
        \def\theequation{\thesection.\arabic{equation}}}

\def\underline#1{\relax\ifmmode\@@underline#1\else
        $\@@underline{\hbox{#1}}$\relax\fi}

\def\titlepage{\@restonecolfalse\if@twocolumn\@restonecoltrue\onecolumn
     \else \newpage \fi \thispagestyle{empty}\c@page\z@
        \def\thefootnote{\fnsymbol{footnote}} }

\def\endtitlepage{\if@restonecol\twocolumn \else \newpage \fi
        \def\thefootnote{\arabic{footnote}}
        \setcounter{footnote}{0}}  

\catcode`@=12
\relax

\def\figcap{\section*{Figure Captions\markboth
        {FIGURECAPTIONS}{FIGURECAPTIONS}}\list
        {Figure \arabic{enumi}:\hfill}{\settowidth\labelwidth{Figure
999:}
        \leftmargin\labelwidth
        \advance\leftmargin\labelsep\usecounter{enumi}}}
 \relax
\def\tablecap{\section*{Table Captions\markboth
        {TABLECAPTIONS}{TABLECAPTIONS}}\list
        {Table \arabic{enumi}:\hfill}{\settowidth\labelwidth{Table
999:}
        \leftmargin\labelwidth
        \advance\leftmargin\labelsep\usecounter{enumi}}}
 \relax
\def\reflist{\section*{References\markboth
        {REFLIST}{REFLIST}}\list
        {[\arabic{enumi}]\hfill}{\settowidth\labelwidth{[999]}
        \leftmargin\labelwidth
        \advance\leftmargin\labelsep\usecounter{enumi}}}
 \relax
\makeatletter
\newcounter{pubctr}
\def\publist{\@ifnextchar[{\@publist}{\@@publist}}
\def\@publist[#1]{\list
        {[\arabic{pubctr}]\hfill}{\settowidth\labelwidth{[999]}
        \leftmargin\labelwidth
        \advance\leftmargin\labelsep
        \@nmbrlisttrue\def\@listctr{pubctr}
        \setcounter{pubctr}{#1}\addtocounter{pubctr}{-1}}}
\def\@@publist{\list
        {[\arabic{pubctr}]\hfill}{\settowidth\labelwidth{[999]}
        \leftmargin\labelwidth
        \advance\leftmargin\labelsep
        \@nmbrlisttrue\def\@listctr{pubctr}}}
 \relax
\makeatother
\newskip\humongous \humongous=0pt plus 1000pt minus 1000pt

\newif\ifdtup

\relax

\def\be{\begin{equation}}
\def\ee{\end{equation}}
\def\ba{\begin{eqnarray}}
\def\ea{\end{eqnarray}}

\def\del{\partial}

\def\a{\alpha}

\def\g{\gamma}
\def\G{\Gamma}
\def\d{\delta}

\def\e{\epsilon}

\def\m{\mu}
\def\n{\nu}
\def\om{\omega}

\def\S{\Sigma}

\def\no{\noindent}

\def\qq{\qquad}

\def\IR{\relax{\rm I\kern-.18em R}}

\def \ha {{1\over 2}}

\def \ov {\over}

\def\IR{\relax{\rm I\kern-.18em R}}
\def\inv{^{\raise.15ex\hbox{${\scriptscriptstyle -}$}\kern-.05em 1}}

\def\tL{{\tilde L}}

\def\wt{\widetilde W}

\def\crbig{\\\noalign{\vspace{3mm}}}

\begin{document}

\renewcommand{\theequation}{\thesection.\arabic{equation}}

\newcommand{\beq}{\begin{equation}}
\newcommand{\eeq}[1]{\label{#1}\end{equation}}
\newcommand{\ber}{\begin{eqnarray}}
\newcommand{\eer}[1]{\label{#1}\end{eqnarray}}
\newcommand{\eqn}[1]{(\ref{#1})}
\begin{titlepage}
\begin{center}

\hfill NEIP-00-014\\
\vskip -.05 cm
\hfill hep--th/0006222\\
\vskip -.05 cm
\hfill June 2000\\

\vskip .6in

{\large \bf BPS solutions and new phases of finite-temperature strings}

\vskip 0.5in

{\bf I. Bakas${}^1$ },\phantom{x}{\bf A. Bilal${}^2$ },\phantom{x} 
{\bf J.-P. Derendinger${}^2$ }\phantom{x}and\phantom{x} 
{\bf K. Sfetsos${}^2$}
\vskip 0.1in
{\em ${}^1\!$Department of Physics, University of Patras \\
GR-26500 Patras, Greece\\
\footnotesize{\tt bakas@ajax.physics.upatras.gr,
bakas@nxth04.cern.ch}}\\
\vskip .2in
{\em ${}^2\!$Institut de Physique, Universit\'e de Neuch\^atel\\
  rue Breguet 1, CH-2000 Neuch\^atel, Switzerland\\
\footnotesize{\tt adel.bilal,jean-pierre.derendinger@unine.ch \\
\vskip -.05in 
sfetsos@mail.cern.ch}}\\
\vskip .2in

\end{center}

\vskip .7in

\centerline{\bf Abstract}

\no
All high-temperature phases of the known $N=4$ superstrings in five
dimensions can be described by the universal thermal potential
of an effective four-dimensional supergravity. This theory,
in addition to three moduli $s, t, u$, contains non-trivial 
winding modes that become massless in certain regions of the thermal 
moduli space, triggering the instabilities at the Hagedorn temperature.
In this context, we look for exact domain wall 
solutions of first order BPS equations. These solutions 
preserve half of the supersymmetries, in contrast to the
usual finite-temperature weak-coupling approximation,
and as such may constitute a new phase of finite-temperature
superstrings.
We present exact solutions for the type-IIA and type-IIB theories and 
for a self-dual hybrid type-II theory. While for the heterotic case the 
general solution cannot be given in closed form, we still present a 
complete picture and a detailed analysis of the behaviour around the
weak 
and strong coupling limits and around certain critical points. 
In all cases these BPS solutions have no instabilities at any
temperature.
Finally, we address the physical meaning of the resulting geometries
within 
the contexts of supergravity and string theory.

\vskip .4 cm
\noindent
\end{titlepage}
\vfill
\eject

\def\baselinestretch{1.2}
\baselineskip 16 pt
\noindent

\def\tT{{\tilde T}}
\def\tg{{\tilde g}}
\def\tL{{\tilde L}}


\section{Introduction}

In the conventional description of string theory, where one has only
very little information about the dynamics of all possible string
states, 
the main framework is provided by an effective supergravity for the 
massless fields which is obtained by integrating out the massive modes.
As a result, many stringy effects that are attributed to the massive
modes cannot be addressed in a systematic way; we can
only appreciate their relevance
in certain regions of the moduli space where massive states can become
massless. In those 
cases, the conventional supergravity approach has to be enlarged
to include the massless as well as all the relevant would-be massless
states on
an equal footing, for otherwise the effective theory will break down due
to
the appearance of singularities. Put differently, the possibility to
have
extra massless states in string theory signals the limitations of the  
perturbative field theory description of string dynamics in all corners
of the moduli space.
Conversely, including the would-be massless fields into the effective
field theory can teach us important non-perturbative lessons about 
string theory.

There are specific interesting problems which can be addressed
systematically, based on supersymmetry, by isolating the dynamics of a 
few relevant modes that can become massless. 
An important example of this kind is provided by the conifold
singularity
in field theory
and its string-theoretical resolution \cite{strom, vo}. In particular,
in
Calabi--Yau compactifications which admit a non-trivial 3-cycle, the
moduli space develops a singularity that invalidates the applicability
of
the conventional low-energy effective theory when the cycle shrinks to
zero size. Incorporating a D-brane that wraps the 3-cycle, with mass
proportional to its period, resolves the problem of conifold
singularities
as the enlarged theory is appropriate for describing cycles of all
sizes.  

Superstrings at finite temperature
provide another interesting example that has been studied on and off for
quite
some time, but admittedly has not been fully explored yet. In
particular, 
it is known that string theory exhibits an exponential
growth of the number of states at high energy, which gives rise to a 
limiting temperature, known as the Hagedorn temperature $T_{\rm H}$ 
\cite{hag}--\cite{alvarez}.
From the world-sheet point of view, 
where one uses a periodic Gaussian
model to describe the propagation of strings on tori, there is a 
Kosterlitz--Thouless phase transition at a critical radius where
vortices 
can condense \cite{sath, ian}, thus leading to a limiting temperature
for string   
thermodynamics (see also \cite{mark}--\cite{alvarez}) 
associated with a phase transition. 
It has been further realized that there are string states with 
non-trivial winding number that can become light close to this
temperature, 
and then turn tachyonic beyond it, thus signaling thermal instabilities 
of string theory at very high temperatures \cite{aw}. Some implications 
of this effect have been examined in the context of strings in the very
early universe, where one tries to find ways that avoid the initial 
singularity, and at the same time explain why the dimensionality of the
physical space-time is four \cite{bv} (see also \cite{neil}, and
\cite{rob} 
for a more recent exposition that takes into account D-branes). 
In an recent development, the 
contribution of the relevant winding modes was explicitly described
in terms of an effective four-dimensional supergravity using a 
universal thermal potential that incorporates 
all phases of $N=4$ superstrings, taking into account the thermal
deformation of the BPS mass formula in $N=4$ supersymmetric theories 
\cite{ak, adk}. However, no solutions  
have been found so far with a concrete physical interpretation and
the ability to   
resolve the problems of quantum cosmology or the subtle issues raised
at the end point of the black-hole 
evaporation; these are situations  
where the temperature grows to infinity and one encounters singularities
in
the context of any perturbative field theory.         

The present paper grew out of the attempt to construct explicit
solutions of
the effective theory of supergravity which was proposed to 
include the dynamics of the lightest  
relevant winding modes that could become massless at the Hagedorn
temperature. 
We construct for the first time families of non-trivial solutions with 
varying winding fields in the type-II and the 
heterotic sectors of string theory. 
We focus attention primarily 
on domain wall solutions, one of the reasons being that they are
technically
the simplest ones to consider; all the fields in the bosonic sector of
the
theory, namely the metric and the collection of scalar fields, are taken
to
depend on a single spatial coordinate only. 
Ultimately, of course, it is 
necessary to search for other types of solutions, 
for instance solutions with spherical symmetry, in order to discuss
models
appropriate for quantum cosmology or black-hole physics. We hope to
report on 
such generalized solutions of the effective supergravity 
elsewhere in the future.   

There are also some other compelling reasons for being interested in the
existence and the explicit construction of domain wall solutions of the 
effective supergravity that describes strings at finite temperature.
The first is provided by the important r\^ole of domain walls, as 
space-time defects, in
a variety of cosmological applications, though we do not focus here on 
these types of physical problems. The second is provided by the fact
that
the domain walls, by their nature, break the four-dimensional Poincar\'e
invariance to an effective three-dimensional 
invariance. Then, one can have a 
working framework for studying the issue of supersymmetry breaking by
non-perturbative effects, as it was done in \cite{ak, adk}, taking
into account the peculiarities of supersymmetry in three dimensions, 
namely that even if local supersymmetry is unbroken the massive
multiplets
will not necessarily exhibit mass degeneracy \cite{3d}.      

The domain wall ansatz allows us to consider solutions of certain 
first order differential equations derived from a 
prepotential. As we explicitly show, these solutions by construction 
preserve half of the supersymmetries, and in this sense they are BPS.
The full set of equations allows consistent truncations to subsectors
which 
correspond to the heterotic string, type-IIA or type-IIB, or the type-II 
string at the self-dual radius (which we call the hybrid type-II). 
For all type-II cases we can find the exact general solutions in closed
form. 
For the heterotic sector, however, no solution in terms of known
functions 
is available, and we instead give a detailed analysis of the solutions 
around the weak and the various strong
coupling limits, as well as around certain critical points, which still
allows us to obtain a reasonably complete overall picture.

It should be stressed that in the standard finite-temperature treatment 
of perturbative superstrings all supersymmetries are broken due to the 
different boundary conditions on bosons and fermions 
one has to impose in the periodic imaginary 
time direction. For this reason, it is not surprising that some modes 
become tachyonic beyond certain temperatures. On the contrary, the 
solutions we find within the domain wall ansatz of the effective 
supergravity are non-pertubative, containing regions of strong coupling, 
and preserve half of the supersymmetries. As such they are expected to 
be stable solutions, and indeed we show that, although the temperature 
can be arbitrarily high, no tachyonic instabilities ever develop.
Even though we are only considering solutions of an effective
supergravity, 
they may well point to a new finite-temperature phase of superstrings 
which is BPS and has no thermal instabilities, i.e. no Hagedorn 
temperature.

To further probe the  physical meaning of these domain wall solutions 
we first study the
propagation of a quantum test particle in the corresponding supergravity 
backgrounds. This allows to discriminate between wave-regular and
non-regular 
geometries, thus narrowing down the physically meaningful set of
solutions.
The analogous question could be addressed within string theory.
The consistency of string thermodynamics seems to require the 
compactness of all spatial directions since otherwise one 
would encounter problems with negative specific
heat (see, for instance, \cite{bv} for a delicate analysis, 
without making reference to the geometrical details of the 
space, and references therein).  It is clear from our explicit solutions 
in the type-II cases, and also true for the heterotic case (with the 
possible exception of certain solutions that are never weakly coupled),
that
none of them is periodic in the spatial dimension that defines the 
domain walls.
However, this does not mean that our solutions of the effective 
supergravity theory are insufficient to extrapolate to string theory. 
Indeed, the compactness criterion is established within the
micro-canonical treatment of string thermodynamics, and this assumes
that 
the theory remains weakly coupled throughout all space.
Furthermore, it concerns the phase where supersymmetry is 
completely broken by the finite temperature. On the other hand, almost
all 
our solutions always contain a 
region of strong coupling.
Most important, our solutions are BPS, preserving half 
of the supersymmetries. These differences are enough to cast serious
doubt on whether the 
arguments about compactness of all spatial dimensions should apply.  
Furthermore, as already noted, the fields corresponding to the winding 
modes of our solutions never become massless even if the temperature 
modulus approaches the would-be Hagedorn temperature, 
which is yet another way to see that these solutions 
belong to a different phase.
  
This paper is organized as follows: In section 2, we review the
main aspects of the effective theory of supergravity that was proposed
to describe the thermal phases of all $N=4$ superstrings and present the
form of the universal thermal potential for the six scalar fields that
correspond to the $s$, $t$, $u$ moduli and the winding states $z_1$, 
$z_2$, $z_3$ that can become massless in the heterotic, type-IIA and
type-IIB
sectors. In section 3, we consider the domain wall ansatz for the metric
and
all the relevant scalar fields. We first discuss quite generally when a 
$D=4$, $N=1$ supergravity theory admits domain wall solutions that are 
derived from first order differential equations. Indeed, this follows 
from some simple reality assumptions about the scalar fields and the 
K\"ahler potential. We show that these solutions automatically preserve 
half (or all) of the supersymmetries and hence are BPS. Then we
specialise 
to the effective thermal supergravity and explicitly derive the coupled 
system of six first order non-linear differential equations for them. 
Consistent truncations of this system lead to various type-II or 
heterotic sectors. In section 4, this system is analysed
explicitly in the type-IIA, type-IIB as well as in the hybrid type-II
sectors where the general solutions can be derived in closed form. 
In section 5, we study the equations
for the heterotic sector which cannot be solved analytically, apart from
a 
very special solution. We are able
to study the behaviour of the solutions in the vicinity of the weak and
the strong coupling regions, as well as around certain critical points,
and present a fairly complete general picture. 
The construction of domain wall solutions
for the whole system of six equations is beyond the scope of the present
paper
and can probably only be done numerically. 
In section 6, we study the r\^ole of boundary conditions in selecting 
physical solutions in the effective theory of supergravity and 
comment on string thermodynamics.  Finally, in section 7, we present our
conclusions and outline some directions for future work.     
In an appendix, we address the issue of having periodic
solutions using criteria from the general theory of dynamical systems.
This may turn out to be useful when studying the complete system
of six equations.   

\vskip 1cm

\section{Thermal potential of effective supergravity}

In this section we present the essential ingredients for constructing 
an effective $N=1$ supergravity in four dimensions\footnote{We always 
count four-dimensional supersymmetries, i.e., $N=1$ supersymmetry 
has four supercharges.} that 
describes the thermal phases of all $N=4$ superstrings in a universal
way, following earlier work \cite{ak, adk, adk2}. The construction 
takes into account the dynamics of the would-be tachyonic winding modes
of the $N=4$ superstrings which are responsible for inducing a phase 
transition at high temperatures in string thermodynamics. 
As such, the effective $N=1$ supergravity theory
provides a systematic framework for quantifying arguments 
about the phase transition occuring at the Hagedorn
temperature due to the special form of the  
effective potential of the relevant winding modes $\omega$, 
which includes a trilinear coupling $\sigma \omega \bar{\omega}$ with
the modulus $\sigma$ describing the temperature \cite{aw}.   

The starting point of the construction is provided by 
five-dimensional $N=4$ theories that are effectively 
four-dimensional at finite temperature. The crucial observation 
is that $d$-dimensional superstrings 
at finite temperature look like $(d-1)$-dimensional strings with 
spontaneously broken supersymmetry. Concretely,
putting strings at finite temperature $T$ amounts to compactifying
the (Euclidean) time on a circle of inverse radius $R^{-1} = 2\pi T$. 
As the Euclidean time direction is compact, one imposes boundary 
conditions to take particle statistics into account; modular invariance
of the thermal partition function dictates then
specific phase factors in the related GSO projection \cite{aw}. 
Technically, this 
procedure is equivalent to a Scherk--Schwarz compactification 
from $d=5$ to $d-1 = 4$ with a well-defined gauging associated to 
the temperature modification of the effective theory of gauged
supergravity \cite{kr, ak}. This gauging of $N=4$ spontaneously 
breaks supersymmetry. 

\subsection{Thermal dyonic modes}

We restrict ourselves to the study of $N=4$ strings, 
because in this case the
supersymmetry algebra and its central extensions suffice to 
determine the masses 
of all BPS states. Recall that the states which become 
tachyonic above a certain Hagedorn temperature $T_{\rm H}$ are
$1/2$-BPS states, at least in dimensions where there are no other BPS
states with smaller fractions of supersymmetry. Then, in $N=4$  
strings, one can identify completely all the perturbative and the  
non-perturbative BPS states that can induce thermal instabilities 
and construct from first principles
an exact effective supergravity for these states, as it was done in 
refs. \cite{ak, adk}. It is useful to observe that the odd dyonic modes 
exhibit the same finite-temperature behaviour as the odd 
winding string states. This follows from the action of dualities on 
the $N=4$ strings, relating 
the dyonic modes to perturbative winding states of dual strings. 
These windings in turn induce thermal instabilities in the dual strings.
Describing the complete set of thermal phases of $N=4$ theories requires
then 
both perturbative and non-perturbative states from all string
viewpoints.

We define the context of the present work, following \cite{ak, adk}, 
by first considering string theories in six dimensions obtained by 
compactification on $T^4$ (heterotic string) or $K3$ (type-II strings). 
We further compactify one dimension $S^1$, with
radius $R_6$, and so the resulting five-dimensional theory exhibits 
T-duality between type-IIA and type-IIB (under $R_6 \rightarrow 
\alpha_{{\rm het}}^\prime/R_6$)
and S-dualities between heterotic and type-II strings. Moreover, since
we are putting strings at finite temperature $T$, the time dimension is 
taken Euclidean and compactified on $S^1$ with radius $R= 1/(2\pi T)$
with 
twisted fermionic boundary conditions that account for the thermal
effects. 

In order to obtain the resulting 
four-dimensional thermal mass formula and examine which states can
become
tachyonic above a certain temperature, thus inducing instabilities, we 
first consider the usual BPS mass formula in $N=4$ supersymmetric
theories
\cite{N=4BPS, duff, cvetic, cardoso, kiritsis} written in heterotic
variables:
\be
\label{BPSmass}
{\cal M}^2 = {1 \over \alpha_{{\rm het}}^{\prime} tu} \Big| m + ntu +
i(m^{\prime} u + n^{\prime} t) + is [\tilde{m} + \tilde{n} tu - i
({\tilde{m}}^{\prime} u + {\tilde{n}}^{\prime} t)] \Big|^2 ~. 
\ee
Here $s$, $t$ and $u$ are defined in terms of the compactification radii
$R_6$, $R$ and the heterotic string coupling $g_{{\rm het}}$ as follows,
\be
s = {1 \over g_{{\rm het}}^2} ~, ~~~~~ t = {RR_6 \over 
{\alpha}_{{\rm het}}^{\prime}} ~, ~~~~~ u = {R \over R_6} ~  ,
\label{tt0}
\ee
supplemented by the relation 
${\alpha}_{{\rm het}}^{\prime} = 4 s$,
in units where the four-dimensional gravitational coupling $\kappa$ has
been normalized to $\sqrt{2}$. These expressions  
will help us later to understand the physical meaning of the various
domain 
wall solutions in terms of the three physical parameters in the problem 
$R_6$, $R$ and $g_{{\rm het}}$.
The integers $m$, $n$, $m^{\prime}$, $n^{\prime}$ are the electric
momentum 
and winding quantum numbers associated to the compactification on the 
2-torus with radii $R_6$ and $R$. The tilded integers are the
corresponding 
magnetic non-perturbative counterparts. Then, string dualities are
simply
described by the interchanges $s \leftrightarrow t$ (heterotic--IIA), 
$s \leftrightarrow u$ (heterotic--IIB) and $t \leftrightarrow u$
(IIA--IIB),
which leave invariant the mass formula and the temperature
radius $R^2 = 4stu$ which is common to all three string theories when
measured in Planck units.      

As discussed in refs. \cite{ak, adk}, the thermal deformation of the
mass formula (\ref{BPSmass}) is very simple: the momentum quantum number
$m$ is replaced by $m + Q^{\prime} + n/2$, $Q^{\prime}$ being the
helicity 
charge. It also reverses the GSO projection, to account for the modified 
boundary conditions in the temperature deformation of the theory. 
For our present purposes, we may restrict our attention to only 
the states having 
$m^{\prime} = n^{\prime} = \tilde{m} = \tilde{n} =0$,
since the states that can become tachyonic first as the temperature
increases
are contained within this subset \cite{adk}.
The thermal spectrum of these light states is then given by 
\be
{\cal M}_T^2 = \left({1 \over R}\left(m + Q^{\prime} + {1 \over 2}kp 
\right) + k \,T_{p,q,r}\, R \right)^2 - 
2 T_{p,q,r} \,{\delta}_{|k| , 1}\,{\delta}_{Q^{\prime} , 0} ~, 
\label{thmass}\ee
where we have defined the integers $k$, $p$, $q$ and $r$ by the
condition
$(n, {\tilde{m}}^{\prime} , {\tilde{n}}^{\prime}) = k (p, q, r)$, 
$p$, $q$ and $r$ being relatively prime numbers, and
$T_{p,q,r}$ denotes the effective string tension 
\be
T_{p,q,r} = 
{p \over {\alpha}_{{\rm het}}^{\prime}} + 
{q \over {\alpha}_{{\rm IIA}}^{\prime}} + 
{r \over {\alpha}_{{\rm IIB}}^{\prime}} ~,  
\ee
which can be written in terms of $s$, $t$ and $u$ using the
identifications
\be
{\alpha}_{{\rm het}}^{\prime} = 4s ~, ~~~~~  
{\alpha}_{{\rm IIA}}^{\prime} = 4t ~, ~~~~~  
{\alpha}_{{\rm IIB}}^{\prime} = 4u ~ ,
\label{tt2}
\ee
again in the Planck units with normalization $\kappa = \sqrt{2}$.   
The mass formula \eqn{thmass} is an extension of the known perturbative
finite-temperature
mass formulas with the correct duality and zero temperature behaviour.
In particular, the last term is generated in perturbative strings by the
GSO 
projection related to the breaking of supersymmetry.
The various constraints imposed among the quantum numbers 
require that $p$, $q$, $r$ are all positive and that $mk \geq -1$.

Analysing the four-dimensional mass formula for ${\cal M}_T^2$, we note 
that tachyons can appear when $Q^{\prime} = 0$ and $|k| = 1$ 
for certain values of the four quantum numbers $m$, $n$, 
${\tilde{m}}^{\prime}$, ${\tilde{n}}^{\prime}$. The critical temperature 
(and hence
the critical radius R) can be found by locating the zeros of ${\cal
M}_T^2$.
The first tachyons and the critical temperatures are: 
\ba
(m, n, {\tilde{m}}^{\prime}, {\tilde{n}}^{\prime}) &=& \pm (-1, 1, 0, 0)
~~~~
{\rm with} ~~ R= (\sqrt{2} + 1) \sqrt{{{\alpha}_{\rm het}^{\prime} \over
2}} 
~, \\ 
(m, n, {\tilde{m}}^{\prime}, {\tilde{n}}^{\prime}) &=& \pm (0, 0, 1, 0)
~~~~
{\rm with} ~~ R= \sqrt{2{\alpha}_{\rm IIA}^{\prime}} 
~, \\ 
(m, n, {\tilde{m}}^{\prime}, {\tilde{n}}^{\prime}) &=& \pm (0, 0, 0, 1)
~~~~
{\rm with} ~~ R= \sqrt{2{\alpha}_{\rm IIB}^{\prime}}.   
\ea
Since the winding numbers in the heterotic, the type-IIA and the type-IIB 
strings are 
respectively $n$, $\tilde m^\prime$ and $\tilde n^\prime$, these three 
pairs of states have winding numbers $\pm 1$, as expected on general 
grounds, in 
their respective perturbative superstring theory. There are also two 
other series of states occuring for $Q^{\prime} = 0$ and $|k| = 1$, with 
quantum numbers $m = -1$ and $(p, q, r)$ arbitrary, which can become
tachyonic 
when $p=1$ or $p=2$; however, in either series the critical temperatures
that
result from ${\cal M}_T^2 = 0$ are higher than  at least one of
the perturbative Hagedorn temperatures of the heterotic, type-IIA or 
type-IIB strings given above.

This analysis of the mass formula suggests the field content to be used 
in the effective field theory description. We certainly need the three 
moduli $s$, $t$ and $u$ and the six scalar fields 
(actually three conjugate pairs of fields)
able to generate the thermal transitions. These states will be embedded 
in the generic scalar manifold of $N=4$ supergravity,
$$
{Sl(2,R)\over U(1)} \,\times\, 
{SO(6,m)\over SO(6)\times SO(m)}\,,
$$
where $m$ is the number of vector multiplets. To study the thermal phase
structure however, it is consistent and sufficient to truncate this 
$N=4$ theory to $N=1$ and to retain only chiral $N=1$ multiplets 
describing the relevant moduli and winding states. After truncation,
the effective theory includes three complex moduli scalars 
$S$, $T$ and $U$,  with scalar manifold
\be
\left({SL(2, R) \over U(1)}\right)_S \times 
\left({SL(2, R) \over U(1)}\right)_T \times 
\left({SL(2, R) \over U(1)}\right)_U  \,,
\ee
and six complex windings $Z^\pm_A$, $A=1,2,3$, living on
\be
\left({SO(2,3) \over SO(2) \times SO(3)}\right)_{Z_A^+} \times
\left({SO(2,3) \over SO(2) \times SO(3)}\right)_{Z_A^-} ~.
\ee
The resulting scalar manifold is a K\"ahler manifold for chiral 
multiplets coupled to $N=1$
supergravity that arises by two successive $Z_2$ projections applied on
the
$N=4$ scalar manifold.  

\subsection{The effective supergravity}

The construction of the effective $N=1$ supergravity Lagrangian 
proceeds as follows \cite{adk}. A generic four-dimensional $N=1$
supergravity
theory is characterized 
by a K\"ahler potential $K$ and a holomorphic superpotential $W$. 
The bosonic sector of the theory is given by the following Lagrangian 
density:
\be
e^{-1}{\cal L} = {1 \over 4}R - {1 \over 2}
K_{I\bar{J}}(\partial^\mu\Phi^I)
(\partial_\mu\bar{\Phi}^{J}) - V(\Phi, \bar{\Phi}) ~,  
\label{sugralagr}
\ee
where $K_{I\bar{J}} = \partial^2 K / \partial \Phi^I \partial
\bar{\Phi}^J$
are the components of the K\"ahler metric for the collection of   
complex scalar fields $\{\Phi^I\}$ present in chiral multiplets. 
The scalar potential $V(\Phi, \bar{\Phi})$ assumes the special form
\be
V= {1\over4}\, e^K \left(K^{I\bar{J}} W_{;I} \overline{W}_{;\bar{J}} - 
3 W \overline{W} \right)\ , 
\label{sugrapot}
\ee
using the notation of covariant derivatives
\be
W_{;I} = {\partial W \over \partial \Phi^I} + {\partial K \over 
\partial \Phi^I} W ~. 
\label{wcovariant}
\ee

The form of the K\"ahler potential $K$ follows from the constraints 
defining the $N=4$ scalar manifold, truncated to $N=1$. One finds, in 
particular,  
\be
\label{Kis}
K = -{\rm ln}(S + \overline{S}) - {\rm ln}(T + \overline{T}) - 
{\rm ln} (U + \overline{U}) 
-{\rm ln}Y_+ - {\rm ln}Y_- ~ ,   
\ee
with 
\be
Y_{\pm} = 1 - 2Z_A^{\pm} {\overline{Z}}_A^{\pm} + (Z_A^{\pm} Z_A^{\pm}) 
({\overline{Z}}_B^{\pm} {\overline{Z}}_B^{\pm}) ~, 
\ee
where summation over repeated indices $A$ or $B=1, 2, 3$ is implicitly 
assumed. The expression of the superpotential $W$ follows from the 
gauging applied to the $N=4$ theory. This gauging encodes the
Scherk--Schwarz supersymmetry breaking mechanism generated by the 
introduction of a finite temperature \cite{ak, adk} and yields   
\be
\label{Wis}
W = 2 \sqrt{2} \left[{1 \over 2} (1 - Z_A^+ Z_A^+)(1 - Z_B^- Z_B^-) + 
(TU -1) Z_1^+ Z_1^- + SU Z_2^+ Z_2^- + ST Z_3^+ Z_3^- \right]\, .  
\ee

The scalar potential of the theory defined by $K$ and $W$ is
complicated,
but focusing on the real directions defined by 
${\rm Im}\, Z_A^{\pm} = {\rm Im}\, S = 
{\rm Im}\, T = {\rm Im}\, U = {\rm Re }\, (Z_A^+-Z_A^-)=0$ leads to 
important simplifications. Furthermore, a study of the mass spectrum in
the
low-temperature limit shows that these directions correspond precisely 
to the possible phase transitions occuring in the
theory. We are then led to consider only the relevant would-be tachyonic
states using the field variables
\be
s = {\rm Re}\,S , ~~~~ t={\rm Re}\,T , ~~~~ u={\rm Re}\,U , ~~~~  
z_A = {\rm Re}\,Z_A^+ = {\rm Re}\,Z_A^- ~. 
\ee
Let us introduce for later convenience the quantities
\be
x^2 = \sum_{A=1}^{3} z_A^2 ~, ~~~~~ H_A = {z_A \over 1 - x^2} ~. 
\label{hreal} 
\ee
Then, in this case, the components of the K\"ahler metric in the field 
space become
\be
K_{S\bar{S}} = {1 \over 4s^2} ~, ~~~ K_{T\bar{T}} = {1 \over 4t^2} ~,
~~~ 
K_{U\bar{U}} = {1 \over 4u^2} ~, ~~~ K_{A^{\pm}{\bar{B}}^{\pm}} = 
{2 \over (1-x^2)^2} {\delta}_A^B ~, 
\ee
i.e., the metric becomes diagonal in the directions $Z_A^{\pm}$; in 
writing the kinetic terms for the fields $z_A$ we have to take 
$K_{AB} = 4 {\delta}_A^B /(1-x^2)^2$, as there is a factor of 2
picked up by $Z_A^+ = Z_A^-$ in this case. 

In order
to present the Lagrangian density for all remaining six scalar fields,
we 
find it helpful to trade $s$, $t$ and $u$ for  
$\phi_1$, $\phi_2$ and $\phi_3$ as follows:
\be
s=e^{-2 \phi_1} ~, ~~~~~ t=e^{-2\phi_2} ~, ~~~~~ u=e^{-2\phi_3} ~.
\ee
Note that $\phi_1 \rightarrow +\infty$ corresponds to the strong
coupling 
limit $s \rightarrow 0$, whereas the weak coupling limit is attained for 
$\phi_1 \rightarrow -\infty$.    
With these definitions, we obtain a simplified form of the 
effective supergravity with bosonic Lagrangian density
\be
e^{-1}{\cal L} = {1 \over 4}R -{1 \over 2} \left(\sum_{i=1}^3
(\partial_\mu\phi_i)^2
+ \sum_{A=1}^{3} {4 \over (1 - x^2)^2} (\partial_\mu z_A)^2 \right) - V
~,  
\label{laagra}
\ee
where the scalar potential is now given by the sum 
\be 
V=V_1+V_2 + V_3\ . \label{dinamiko} 
\ee 
Each term $V_A$ is a function of the moduli $\phi_i$ and a polynomial
in $H_A$. Explicitly, we have 
\ba 
V_1& = &
{1 \over 2}e^{2\phi_1} H_1^2 \left({\rm cosh}(2 \phi_2+2 \phi_3)(4
H_1^2+ 
1)- 3  
\right)\ ,\\ 
V_2& = & {1 \over 4}e^{2\phi_2} H_2^2
\left(e^{-2\phi_1-2\phi_3}(4H_2^2+1) -4 \right)\ ,\\ 
V_3& =& {1 \over 4}  
e^{2\phi_3} H_3^2 \left(e^{-2\phi_1-2\phi_2}(4H_3^2+1) -4 \right)\ .
\ea
Our normalizations have been chosen so that the kinetic terms of the 
fields $\phi_i$ assume their canonical form, with an overall factor
$1/2$ 
as coefficient, since the normalization $\kappa=\sqrt2$ has been used.
Then, the scalar potential $V$ provides the universal thermal effective 
potential that describes all possible high-temperature instabilities of
$N=4$ 
superstrings.  

At this point, it is very useful to observe that the scalar potential 
(\ref{dinamiko}) for the six real fields $s$, $t$, $u$ and $z_A$ 
can also be cast into the special form 
\be
V={1 \over 4} \left( \sum_{i=1}^3 \left({\partial \wt \over 
\partial \phi_i}\right)^2 + {1 \over 4} \sum_{A=1}^3 (1 - x^2)^2 
\left({\partial \wt \over \partial z_A}\right)^2 - 
3 {\wt}^2 \right) \ .
\label{dina1}
\ee
This property only depends on the structure of the K\"ahler potential
$K$.
The `prepotential' $\wt$ given by 
\be 
\begin{array}{rcl}
\wt &=& {1 \over 2} 
e^{\phi_1+\phi_2+\phi_3} - 2 e^{\phi_1} {\rm sinh} (\phi_2+\phi_3) 
H_1^2 +e^{-\phi_1}\left(e^{\phi_2-\phi_3} H_2^2 + e^{\phi_3-\phi_2}
H_3^2
\right) \crbig
&=& {1\over\sqrt{stu}}\left( {1\over2} + (tu-1)H_1^2 + suH_2^2 + stH_3^2 
\right)\,.
\end{array}
\label{supereal} 
\ee 
The second expression indicates that the prepotential is actually given
by
\be
\wt = e^{K/2} W \mid_{{\rm real ~  directions}} ~,  
\ee
as can be inferred from the general form of the supergravity potential
and from our specific K\"ahler potential\footnote{In particular, 
${\rm exp}(-K/2) = 2 \sqrt{2 stu}\ (1-x^2)^2$ upon restriction to the 
real directions of the theory.} (\ref{Kis}) and superpotential
(\ref{Wis}). As we will see in section 3.2, this is not an accident.
The prepotential will be of particular importance when
writing the system of first order differential equations for
the domain wall configurations of the theory.  
It should be emphasized that given the potential \eqn{dinamiko},  
we may view \eqn{dina1} as a differential equation for the prepotential 
$\wt$. Hence,  in general, there can be solutions other than
\eqn{supereal}. However, the latter choice is the one that 
leads to one-half supersymmetric solutions, as we will see in 
section 3.3. In general,  another choice for $\wt$ satisfying 
\eqn{dina1} would break supersymmetry completely.

\subsection{Thermal phases of $N=4$ strings}

The thermal phase structure of $N=4$ superstrings can
be deduced from the form of the potential $V$. First of all note that 
the low-temperature phase, i.e. large $R$, or large $stu$, 
or large and negative $\phi_1+\phi_2+\phi_3$, 
is common to all strings. It is characterized
by the condition $H_1 = H_2 = H_3 = 0$, in which case $V_1 = V_2 = V_3 =
0$,
and so the potential vanishes for all values of the moduli fields $s$, 
$t$ and $u$. Since the four-dimensional couplings of the three strings
are
\footnote{Notice that there is a factor of $\sqrt{2}$ that rescales
the right hand sides of \eqn{tt1} as compared with the corresponding
equation 
for zero temperature \eqn{tt0}. This is due to the new spin structure
introduced
by temperature. However, eq. \eqn{tt2} for the 
three string scales remains the same.}
\be
s = {\sqrt{2} \over g_{\rm het}^2} ~, ~~~~~ 
t = {\sqrt{2} \over g_{\rm IIA}^2} =\sqrt{2} {R R_6\ov \a'_{\rm het}}~,
~~~~~ 
u = {\sqrt{2} \over g_{\rm IIB}^2} = \sqrt{2} {R\ov R_6}~ ,
\label{tt1}
\ee
we see that this phase exists in the perturbative regime of each string
theory.
Notice also that the configuration $H_A=0$ and $\phi_i$ constant (but 
arbitrary) is an 
exact solution of the theory, with {\it all 
supersymmetries broken} by the thermal deformation. 
The reason is that in this phase the supersymmetry transformation 
is proportional to 
\be
e^{K/2}\,W_{;I} = {1\over2\sqrt{stu}}{\partial K\over \partial\Phi_I} \
,
\ee
which does not vanish in the moduli directions, except of course 
in the limit $stu\rightarrow \infty$ (zero-temperature limit). 

There are three high-temperature phases generated by each of the three 
contributions to the scalar potential (\ref{dinamiko}), and which are  
characterized by a non-zero value of a winding state $z_A$. Which phase
is 
selected depends on the values of the moduli $\phi_i$, corresponding to
the
coupling, the temperature radius $R$ and the sixth-dimension radius
$R_6$
in each string theory (with string units $\alpha^\prime_{\rm het}$, 
$\alpha^\prime_{\rm IIA}$ or $\alpha^\prime_{\rm IIB}$). None of them is 
a solution to the theory with constant scalars. The detailed analysis of 
the thermal phase structure \cite{adk} leads to the following 
classification.

\no\underline{Heterotic high-T phase}: \\
In heterotic-string 
units, 
the temperature modulus is $tu=1/(2\pi^2 \a'_{\rm het}T^2)$, while it is
$stu=1/(2\pi \kappa T)^2$ in Planck units. 
A heterotic tachyon $H_1$ appears first if
\be
\label{highhet}
(\sqrt{2}-1)^2 <tu <(\sqrt{2}+1)^2 ~, \qq su > 4\ , \qq st > 4\  ,
\ee
where the last two condition guarantee the absence of type-II tachyons.
Then, this implies for the four-dimensional heterotic coupling that 
\be
\sqrt2 s^{-1} = g_{\rm het}^2 <  g_{\rm crit}^2 = {\sqrt{2} + 1 \over 
2\sqrt2} ~. 
\ee
The upper and lower critical temperatures in \eqn{highhet} are related
by T-duality in the temperature radius.
Within the interval (\ref{highhet}), the theory has a (non-tachyonic)
solution 
for $H_1=1/2$, $tu=1$. And, since the scalar potential reduces to 
\be 
V = -{1\over8}e^{2\phi_1}\ , 
\ee
a linear dilaton background is a solution. This background breaks 
one half of the supersymmetries \cite{adk}. This particular solution
will
be rederived as a special case of a more general class of domain wall
backgrounds that break half of the supersymmetry in the heterotic
sector, 
in section 5.  

In the strong coupling regime of the heterotic theory, where 
$s\rightarrow 0$, there are only type-II instabilities; the 
high-temperature heterotic phase cannot be reached
for any value of the radius $R_6$ if $g_{\rm het}^2 > g_{\rm crit}^2$.   
Put differently, the high-temperature heterotic phase can only exist in 
the perturbative heterotic regime, which by the heterotic--type-II
duality
corresponds to the non-perturbative regime of the type-II string theory.

\no\underline{Type-IIA/B high-T phases}: \\
The relevant temperature modulus is either $su=1/(2\pi^2 \a'_{\rm
IIA}T^2)$ 
for type-IIA or $st=1/(2\pi^2 \a'_{\rm IIB}T^2)$ for type-IIB strings.
The high-temperature phase for type-II strings is then defined by 
$su<4$ (for IIA) 
or $st<4$ (for IIB). Thus, type-II
instabilities arise in the strong coupling region of the heterotic
string. 
The analysis of the problem shows that the value 
$R_6 /\sqrt{{\alpha}_{\rm het}^{\prime}}$ determines whether a type-IIA 
or a IIB tachyon comes first as the temperature increases. We have in
particular the following two high-temperature type-II phases:
\ba
{\rm IIA} & : &\qq 
2\pi T > {1\over2 g^2_{\rm het}} {1\over R_6}
~, ~~~~ g_{\rm het}^2 > {\sqrt{2} + 1 \over 2\sqrt2} ~, ~~~~ R_6 <
\sqrt{{\alpha}_{\rm het}^{\prime}} ~, \\ 
{\rm IIB} & : & \qq
2\pi T >  {1\over2g^2_{\rm het}} {R_6\over\alpha^\prime_{\rm het}}
~, ~~~~ g_{\rm het}^2 > {\sqrt{2} + 1 \over 2\sqrt2} ~, ~~~~ R_6 > 
\sqrt{{\alpha}_{\rm het}^{\prime}} .~
\ea

Note that the type-II high-temperature phases differ from the heterotic
one
by the absence of an exact solution with constant windings and/or
moduli. 
In these phases, the theory depends on a non-zero, field-dependent
winding
and two non-trivial moduli, $t$ and $su$ in the IIA case, $u$ and $st$
for
IIB strings. 
This concludes our general review of the thermal aspects of $N=4$
strings at
finite temperature. 

\vskip 2.cm

\section{BPS domain wall solutions}
\setcounter{equation}{0}

We are now in the position to look for solutions of the effective
supergravity that describe the thermal phases of $N=4$ superstrings.
We will use the domain wall ansatz for two reasons.
First, this  probably is the simplest possible class of 
solutions one may construct in supergravity theories whose bosonic
sector is described by gravity coupled to a selection of scalar fields
with non-trivial potential terms, as the universal thermal effective
potential we have at our disposal. Second, with all fields depending 
only on a single variable one can look for solutions that satisfy more
stringent, but easier to solve, first order differential equations
rather 
than second order ones. We will show quite generally that all solutions 
of these first order equations preserve (at least) half of the 
supersymmetries and hence are BPS.

We will first discuss in a general setting when 
gravity coupled to scalars admits solutions of 
first order equations. Then we will 
see under which curcumstances this is true for the bosonic sector of 
supergravity, and that in these cases the solutions are BPS. Finally,  
we specialise to the case of present interest and derive the set of 
coupled first order equations of the effective supergravity describing 
the thermal phases of  $N=4$ superstrings.

\subsection{First order equations for gravity coupled to scalars}

\setcounter{equation}{0}

Consider the following $D$-dimensional Lagrangian density representing
gravity coupled to $N$ scalars 
\be
{1\over \sqrt{g}}\ {\cal L} 
= {1\ov 4} R - \ha  G_{IJ}(\varphi) g^{mn} \del_m \varphi^I
\del_n \varphi^J - V(\varphi) \ ,
\ee
where summation over the repeated indices $I,J=1,2,\dots, N$  and 
$m,n = 1,2,\dots , D$ is implied. The equations of motion that follow
from varying 
the metric and the scalar fields are:
\ba
\d g & : & \qq {1\ov 4} R_{mn} -\ha G_{IJ} \del_m \varphi^I \del_n
\varphi^J
={1\ov D-2} g_{mn} V\ , \\
\d \varphi & : & \qq D^2 \varphi^I 
+ \G^I_{JK} g^{mn}\del_m \varphi^J \del_n \varphi^K=
G^{IJ} \del_J V\ ,
\ea
where $\G^I_{JK}$ is the Christoffel symbol formed using the
scalar-field 
space metric $G_{IJ}$.
Let us assume that the potential $V(\varphi)$ is given in terms of a
prepotential 
$\wt(\varphi)$ as:
\be
V= {\beta^2 \ov 8} \left( G^{IJ} \del_I \wt \del_J \wt 
-2 {D-1\ov D-2} \wt^2\right)\ ,
\label{vwt}
\ee
where $\beta$ is a dimensionful constant.

Let us make the ansatz for a metric of the form 
\be
ds^2 = dr^2 + e^{2 A(r)} \eta_{\m\n} dx^\m dx^\n\ , 
\qq \m,\n=1,2,\dots,D-1\ ,
\label{mee2}
\ee
that preserves the flat space-time symmetries in
$D-1$ dimensions. We also assume that all scalars depend only on the
variable
$r$, i.e., $\varphi^I = \varphi^I(r)$. 
Then, it can be shown by a straightforward calculation that if the
following
first order 
equations hold 
\be {d\varphi^I\ov dr} = \pm {\beta\ov 2} G^{IJ} \del_J \wt\ ,
\qq {dA\ov dr} =\mp {\beta\ov D-2} \wt\ ~, 
\label{fffoe}
\ee
the second order equations are also satisfied.\footnote{Recently, 
such first order equations with flat metric in the scalar-field space
appeared in the context of supersymmetric solutions
in five-dimensional gauged supergravity 
(for a first example, see \cite{FGPW1}).
For cases where the metric in the scalar-field space is non-trivial,
five-dimensional 
examples have been given in \cite{WFGK}.}
In proving this statement we use the fact that the non-vanishing 
components of the Riemann tensor for the metric \eqn{mee2} are
\ba 
R_{\m\n} & = &  -\eta_{\m\n} e^{2A}\left(A^{\prime\prime} + (D-1) 
(A^{\prime})^2 \right)\ ,
\nonumber\\
R_{rr} & = &  - (D-1) \left(A^{\prime\prime} +  (A^{\prime})^2\right)\ ,
\ea
where the prime denotes the derivative with respect to $r$. 
Note that the freedom in the choice of sign in \eqn{fffoe} is due to the 
fact that \eqn{vwt} does not determine the sign of $\wt$. Equivalently 
it expresses the freedom to change $r\to -r$.

Throughout the remainder of this paper we restrict to  $D=4$ and 
choose $\beta=\sqrt{2}$ (consistent with $\kappa=\sqrt{2}$).

\subsection{First order equations for the bosonic sector of
supergravity}

Consider now the bosonic sector of a generic four-dimensional $N=1$ 
supergravity
theory as  given by eqs. \eqn{sugralagr}, \eqn{sugrapot} and
\eqn{wcovariant}.
Assume that all ${\rm Im}(\Phi^I)$ vanish and that the K\"ahler
potential 
is such that
\be
2\left.{\partial K\over\partial\Phi_I}\right|_{\rm real\,\,directions}
= {\partial\over\partial{\rm Re}\,\Phi_I}\,\left. \Bigl(K
\right|_{\rm real\,\,directions}\Bigr)\ .
\label{kassumption}\ee
This is indeed the case for our thermal supergravity, 
but for the time being we do not want to consider any specific model.
The definition of the covariant derivative \eqn{wcovariant}
of the superpotential $W$ and the assumption \eqn{kassumption} imply
that
if we define
\be
\wt =\left. e^{K/2}W\right|_{\rm real\,\,directions}\ 
\ee
we will have
\be
{\partial\over\partial\Phi^I}\wt = e^{K/2}W_{;I}\ .
\label{wtvw}
\ee
As a consequence, the scalar potential \eqn{sugrapot}  of the 
supergravity can be rewritten using only $\wt$ as
\be
V={1\over 4} \left( K^{I\overline J}{\partial\wt\over\partial\Phi^I}
{\partial\wt\over\partial\Phi^J} - 3 \wt^2 \right)\ .
\label{potwt}
\ee
We see that this is precisely of the form \eqn{vwt} with $D=4$ and 
$\beta=\sqrt{2}$.

We conclude that if we find solutions to the domain wall ansatz
\eqn{mee2}
satisfying the first order equations
\be {d\Phi^I\ov dr} = \pm {1\ov \sqrt{2}} K^{I\overline J} 
{\partial\wt\over\partial\Phi^J}\ ,
\qq {dA\ov dr} =\mp {1\over \sqrt{2}} \wt\ ~, 
\label{sugrafoe}
\ee
and take the fermions to vanish, 
then all equations of motion of the supergravity will be satisfied.
We now proceed to show that these solutions of \eqn{sugrafoe} precisely 
preserve half or all of the four-dimensional supersymmetries.

\subsection{Supersymmetry of the solution}\label{secsusy}

It is very simple to obtain the number of supersymmetries left unbroken
by a solution of the first order equations. The only additional
assumptions we need is that the K\"ahler potential satisfies
\eqn{kassumption},
${\rm Im}(\Phi^I)$ vanishes, and
that the fields only depend on a single coordinate, say $r$, with
$g^{rr}=1$.

Consider then the supersymmetry transformation of the (left-handed)
fermionic component $\chi^I_L$ of a chiral multiplet with scalar
$\Phi^I$:
\be
\delta\chi^I_L = {1\over\sqrt2}\gamma^\mu\,
(\partial_\mu\Phi^I)\epsilon_R -
{1\over2}\epsilon_L\,e^{K/2} K^{I\overline J}W_{; J} 
+ {\rm fermionic\,\,terms}\ .
\ee
Since, by assumption, the solution depends only on a single coordinate
$r$ 
with metric 
$g^{rr}=1$, we have 
$\gamma^\mu \partial_\mu\Phi^I=\gamma^r  \partial_r\Phi^I$ where
$\gamma^r$ 
satisfies $(\gamma^r)^2=1$. Then, using the first order equation  
\eqn{sugrafoe} (and \eqn{wtvw}) we have
\be
\gamma^\mu\partial_\mu\Phi^I = \pm
{1\over\sqrt2}\gamma^r K^{I\overline J}{\partial\wt\over\partial\Phi^J}
= \pm {1\over\sqrt2}\gamma^r e^{K/2}K^{I\overline J} W_{;J}\ .
\ee
It follows that for vanishing fermions
\be
\delta\chi^I_L =-{1\over2}P_L(1\mp  \gamma^r)\epsilon \ 
e^{K/2}K^{I\overline J}W_{;J}\ ,
\ee
where $P_L$ is the left-handed chirality projector (of course there is a
similar equation for $\chi^I_R$). Clearly, since $(\gamma^r)^2=1$, 
the solution will preserve one half of the supersymmetries if $W_{;J}\ne
0$, 
and all supersymmetries if $W_{;J}=0$, as it should.

In order to compute the dependence of the Killing spinor $\e$ 
on the space-time coordinates, 
one in principle has to solve the gravitino equation. Equivalently,
for static configurations (as in our domain wall ansatz \eqn{metriki})
one can 
use an argument based on the supersymmetry algebra and deduce that 
$\e=g_{00}^{1/4}\e_0$, where $\e_0$ is a constant spinor subject to 
model-dependent 
projection conditions, which reduce the number of its independent 
components \cite{kallosh}. Applying this to our case we find 
\be
\e(r) = e^{A(r)/2} \e_0\ ,\qq (1 \mp \g^r)\e_0= 0\ .
\ee
This result should be interpreted with some care when considering 
strings at finite temperature. Formulating a theory at finite 
temperature requires a rotation to Euclidean compact time and, 
with a trivial background, supersymmetry is completely broken. 
As already discussed in section 2.3, this is the case
of the low-temperature phase. 
Our domain wall solutions are, however, 
backgrounds depending on a spacelike variable $r$ with 
a manifest three-dimensional 
residual ``space-time" symmetry, with signature $(3,0)$ at finite 
temperature. The supercharges preserved by these backgrounds
are then supersymmetries inherited from the original five-dimensional 
supersymmetry algebra, with Minkowski signature $(4,1)$.

We conclude that the solutions of the first order equations
\eqn{sugrafoe}
are 1/2 BPS solutions, except of course if 
${\partial\wt\over\partial\Phi^I}=0$, in which case the scalar fields
are 
constant and supersymmetry is unbroken.

\subsection{The first order equations for the thermal supergravity}

Next, we specialise the discussion to the supergravity theory that 
describes the thermal phases of the $N=4$ superstrings, 
by applying the 
restrictions and identifications explained in section 2.2. The K\"ahler 
metric, the potential $V$
and prepotential $\wt$ are given in eqs. \eqn{laagra}, \eqn{dinamiko}
and 
\eqn{supereal}. There are six real scalar fields and the K\"ahler metric
is 
diagonal $K_{ij}=K_{(i)} \delta_{ij}$. For convenience we write the 
relevant equations again, for $D=4$. The metric is assumed to be
conformally flat and written in the form (see, for instance, \cite{cvso})
\be
ds^2 = dr^2 + e^{2A(r)} \eta_{\mu \nu} dx^{\mu} dx^{\nu} ~,\qq  \mu,\n
=1,2,3\ 
\label{metriki}
\ee
and the BPS domain wall configurations are solutions 
of the following system of
first order differential equations:
\be
{d \varphi_i \over dr} = - {1 \over \sqrt{2}} K_{(i)}^{-1} 
{\partial \wt \over \partial \varphi_i} ~; \qq
{dA \over dr} = {1 \over \sqrt{2}} \wt \ ,
\label{Weqs}
\ee
taking into account the inverse of the diagonal metric in field space 
for each one of the components $\varphi_i$.
Note that we  have used the freedom $r\to -r$ to make a definite choice 
of signs in these equations.

The six scalar fields are naturally separated into two groups:
$\{\phi_i ; ~ i = 1, 2, 3 \}$ corresponding to $s$, $t$, $u$ and the
three
winding fields
$\{z_i ; ~ i = 1, 2, 3 \}$. For 
the first set the metric in field space is flat, i.e., 
$K_{(i)}(\phi) = 1$, whereas for the second set all components are
equal but non-trivial, namely $K_{(i)}(z) = 4/(1-z_1^2 -z_2^2
-z_3^2)^2$,
see eq. \eqn{laagra}. The prepotential $\wt$ was given in
\eqn{supereal}.
As a result, the domain walls of the theory correspond
to solutions of the non-linear system
\ba 
\sqrt{2} {d \phi_1 \over dr} & = &  
-{1\over 2} e^{\phi_1+\phi_+} + 2 e^{\phi_1} \sinh\phi_+\, H_1^2 
+e^{-\phi_1} \left(e^{\phi_-} H_2^2 + e^{-\phi_-} H_3^2 \right) 
\, ,
\label{fi1}\\
\sqrt{2} {d \phi_2 \over dr} & = &  
-{1\over 2} e^{\phi_1+\phi_+} + 2 e^{\phi_1} \cosh\phi_+\,  H_1^2  
-e^{-\phi_1} \left(e^{\phi_-} H_2^2 - e^{-\phi_-} H_3^2 \right) 
\, ,
\label{fi2}\\
\sqrt{2} {d \phi_3 \over dr} & = &  
-{1\over 2} e^{\phi_1+\phi_+} + 2 e^{\phi_1} \cosh\phi_+\,  H_1^2  
+e^{-\phi_1} \left(e^{\phi_-} H_2^2 - e^{-\phi_-} H_3^2 \right) 
\, ,
\label{fi3}\\
\sqrt{2} {d \ln H_1\over dr}& = &   
e^{\phi_1}\sinh\phi_+\,  ( 4H_1^2+1)
-2e^{-\phi_1} \left(e^{\phi_-} H_2^2 + e^{-\phi_-} H_3^2 \right) 
\, ,
\label{hh1}\\
\sqrt{2} {d \ln H_2\over dr}& = &  
-{1\over 2} e^{-\phi_1+\phi_-} 
+4 e^{\phi_1}\sinh\phi_+\,  H_1^2
-2e^{-\phi_1} \left(e^{\phi_-} H_2^2 + e^{-\phi_-} H_3^2 \right) 
\, ,
\label{hh2}\\
\sqrt{2} {d \ln H_3\over dr}& = &    
-{1\over 2} e^{-\phi_1-\phi_-} 
+4 e^{\phi_1}\sinh\phi_+\,  H_1^2
-2e^{-\phi_1} \left(e^{\phi_-} H_2^2 + e^{-\phi_-} H_3^2 \right) 
\, ,
\label{hh3} 
\ea
where we found it more convenient to work with the fields 
$H_i = z_i /(1 - z_1^2 - z_2^2 - z_3^2)$ instead of $z_i$
and we defined $\phi_\pm = \phi_2 \pm \phi_3$. 
As soon as a solution has been found, the conformal factor of the 
metric can be obtained by a simple integration of the resulting 
expression for the prepotential $\wt$, as it has been prescribed above.

Although these equations are first order, their explicit solution 
is still a difficult task 
because they are highly non-linear in the 
general case. We note that there are examples in the scalar sector
of maximal gauged supergravity in 4, 5 and 7 space-time dimensions
where the resulting system of equations can be linearized by 
introducing appropriate combinations of fields plus an algebraic
constraint. This led to a systematic classification of the 
possible domain wall solutions in terms of Riemann surfaces with 
varying genus (for details, see the method presented in \cite{bs, bbs}). 
However, things do not always work that way; as it turns out the 
above system of equations that describe strings at 
finite temperature can only be partially studied by similar 
methods.

It can be easily seen from the last three equations that
the fields $H_i$ can be solely expressed in terms of the fields
$\phi_i$, but one should avoid substituting their integral expressions
into the first three equations because it would lead to a complicated
system of integro-differential equations among the fields $\phi_i$
alone. In either case, the system of equations at hand is rather
complicated and difficult to solve in general.
Instead, we will focus attention on subsectors obtained by
consistent truncations of the field content. 
Each one of these consistent truncations  
results into a system of three first order equations that  
correspond to the various type-II and heterotic theories. Luckily, it
will 
turn out that the general solutions 
for all type-II theories can be found explicitly. However,
in the heterotic case a general solution cannot be given in closed form.
Hence in this case, apart from extracting 
the behaviour of the fields in the strong and weak coupling regions 
as well as around certain critical points, one has 
unavoidably to rely on numerical studies. Combining these 
pieces of information,
we can nevertheless give a rather complete picture of how the different 
solutions behave.

\vskip 2.cm

\section{Type II string theories}
\setcounter{equation}{0}

The simplest truncations of
the domain wall equations lead to different type-II sectors. 
We will first examine the type-IIA and IIB theories, which can be
treated simultaneously using the appropriate field identifications, 
and then examine a hybrid type-II sector, which describes type-II
strings
at the self-dual radius, and turns out to be exactly solvable, too.
   
\subsection{Type-IIA and IIB sector}
According to the field identifications made in the effective
supergravity,
the type-IIA and type-IIB sectors of the theory are obtained by setting
\ba
&& z_1 =  z_2 = 0 ~, ~~~~~ {\rm for ~ type\ IIB} \\
&& z_1 =  z_3 = 0 ~, ~~~~~ {\rm for ~ type\ IIA} 
\ea
in which case $H_1 = H_2 = 0$ and $H_1 = H_3 = 0$ respectively. It
follows 
from \eqn{fi1} and \eqn{fi2} or \eqn{fi3} that $\phi_1$ equals $\phi_2$ 
or $\phi_3$, respectively,
up to an irrelevant additive constant which we will ignore.
Then, it is convenient to set 
\be
\phi_1 = \phi_2 ={\phi\ov \sqrt{2}} ~; ~~~~ \phi_3 = \chi ~; ~~~~ 
z_3 = z ~~~~~ {\rm for ~ IIB}  \ ,
\ee
or 
\be
\phi_1 = \phi_3 = {\phi\ov \sqrt{2}} ~; ~~~~ \phi_2 = \chi ~; ~~~~ 
z_2 = z ~~~~~ {\rm for ~ IIA} 
\ee
and introduce in either case the field $\omega$ for the winding field
\be
z = {\rm tanh}\left({\omega \over 2}\right) \ ,
\ee
for which we have $2H = {\rm sinh} \omega$. 
The appropriate string coupling is $g_{\rm II}\sim e^\chi$ and
the temperature field 
is $T\sim e^{\sqrt{2} \phi}$, when it is normalized
in type-II string units that remove the $\chi$-dependence.
We treat both cases together because the (pre)potential assumes the
same form for IIA and IIB, and the same is true for the truncated
system of differential equations for the domain walls. Hence, no
distinction will be made in the following between type-IIA or type-IIB.
Note also that the kinetic terms for the fields $\chi$, $\phi$ and $\om$
all
assume their canonical form.  

Explicit calculation shows that in terms of the new variables the
prepotential \eqn{supereal} becomes
\be
\wt_{{\rm II}} = {1 \over 2} e^{\chi} \left(e^{\sqrt{2} \phi} + 
{1 \over 2} e^{-\sqrt{2} \phi} {\rm sinh}^2 \omega \right) \ , 
\ee
whereas the corresponding potential \eqn{dinamiko} is
\be
V_{{\rm II}} = {1 \over 16} e^{2 \chi} \sinh^2\om\left(e^{-2 \sqrt{2}
\phi} 
\cosh^2 \omega - 4 \right)\ .
\ee
It is clear from this potential $V_{\rm II}$ that $\om$ becomes
tachyonic if
$e^{-2 \sqrt{2} \phi}<4$ at $\om=0$, in accordance with the discussion 
of section 2.3. Using the truncated prepotential $\wt_{\rm II}$,
the type-II domain walls obey the simpler system of first order
equations 
\ba
{d \chi \over dr} & = & -{1 \over 2 \sqrt{2}} e^{\chi} \left(
e^{\sqrt{2} \phi} + {1 \over 2} e^{-\sqrt{2}\phi}
{\rm sinh}^2 \omega \right) ,
\nonumber\\
{d \phi \over dr} & = & -{1 \over 2} e^{\chi} \left(e^{\sqrt{2}\phi} 
- {1 \over 2} e^{-\sqrt{2}\phi} {\rm sinh}^2 \omega \right) , 
\label{syII}\\
{d \omega \over dr} & = & -{1 \over 4\sqrt{2}} e^{\chi-\sqrt{2}\phi}
\sinh 2\omega ~.
\nonumber
\ea
These equations can as well be obtained from the full set of 
first order equations \eqn{fi1}-\eqn{hh3} by imposing the above field 
identifications for the type-IIA or IIB strings.
The equation for the conformal factor of the metric \eqn{metriki} is
easily 
integrated and gives $A=-\chi$ (up to a constant that can be absorbed
into
a redefinition of the $x^\m$'s). The system of equations \eqn{syII} is 
invariant under $\om\to -\om$. Therefore we will consider the case 
$\om\ge 0$ without any loss of generality. 

We note for completeness that in the type-II sector one may fully absorb 
the $\chi$-dependence that appears on the right hand side of these
differential equations by simply changing the coordinate variable
$r$ to $\zeta$ defined as $d\zeta = e^\chi dr$. Put 
differently, $\chi$ plays the r\^ole of a Liouville field in that the
``gravitational dressing" of an auxiliary massive model made out 
of the fields  
$\phi$ and $\omega$ 
\be
{\cal L}(\phi, \om) = -{1 \over 2} (\partial \phi)^2 - {1 \over 2} 
(\partial \omega)^2 - {1 \over 16} \sinh^2\om\left(e^{-2 \sqrt{2} \phi} 
\cosh^2 \omega - 4 \right)\ ,
\ee
yields the scalar field sector of the truncated type-II theory with 
all three fields $\phi$, $\omega$ and $\chi$ having kinetic terms in
canonical form and a dressed potential that assumes the form 
$V_{{\rm II}}$ above.\footnote{This interpretation is motivated by
two-dimensional gravity coupled to massive $\sigma$-models. However,
unlike other examples, where non-abelian Toda theories result in this
fashion \cite{ioa}, the present model does not have a special 
meaning in integrable systems.} 
 
It is an interesting finding that the domain wall equations can be
completely integrated for the type-II sector, and one has for the
first time a one-parameter family  of explicit solutions with
non-trivial
winding $\omega$. 
It is convenient for this purpose to present the 
solution by treating the field $\om$ as an independent variable,
which is legitimate since the third equation in \eqn{syII} implies that 
$\om$ is a monotonous function of $r$.
Then, the differential equation for $\phi$ can be easily integrated
and also the equation for $\chi$.
The final result is a family of BPS solutions, as shown in section 3.3, 
with
\ba
e^{-2\sqrt{2} \phi} & = & 2 \cosh^2\om\ (\ln \coth^2\om +c) -2\ ,
\nonumber \\
e^{2 \chi} & = &\cosh^2 \om\ e^{\sqrt{2}\phi}\ ,
\label{chiII}
\ea
parametrized by an arbitrary integration constant $c$.
As we will see, the physical interpretation of the solutions  
depends crucially on whether $c$ is positive, negative or zero.
The behaviour of the functions $e^{2\sqrt{2}\phi}$ and $e^{2 \chi}$ for
the 
three choices of $c$ is sketched in Figure 1.
We also note that we have omitted a
multiplicative integration constant on the right hand side of the 
second equation in \eqn{chiII}, 
since it can always be absorbed into trivial field redefinitions.
We see that, independently of $c$, the behaviour of 
$T^{-2}\sim e^{-2 \sqrt{2} \phi} $ as $\om\to 0$ is 
$e^{-2 \sqrt{2} \phi} \simeq 2 \ln \om^{-2}\to \infty$. Hence,
the winding mode cannot be tachyonic for all these solutions.

\begin{figure}[ht]
\begin{center}
{\scalebox{.6}{\includegraphics{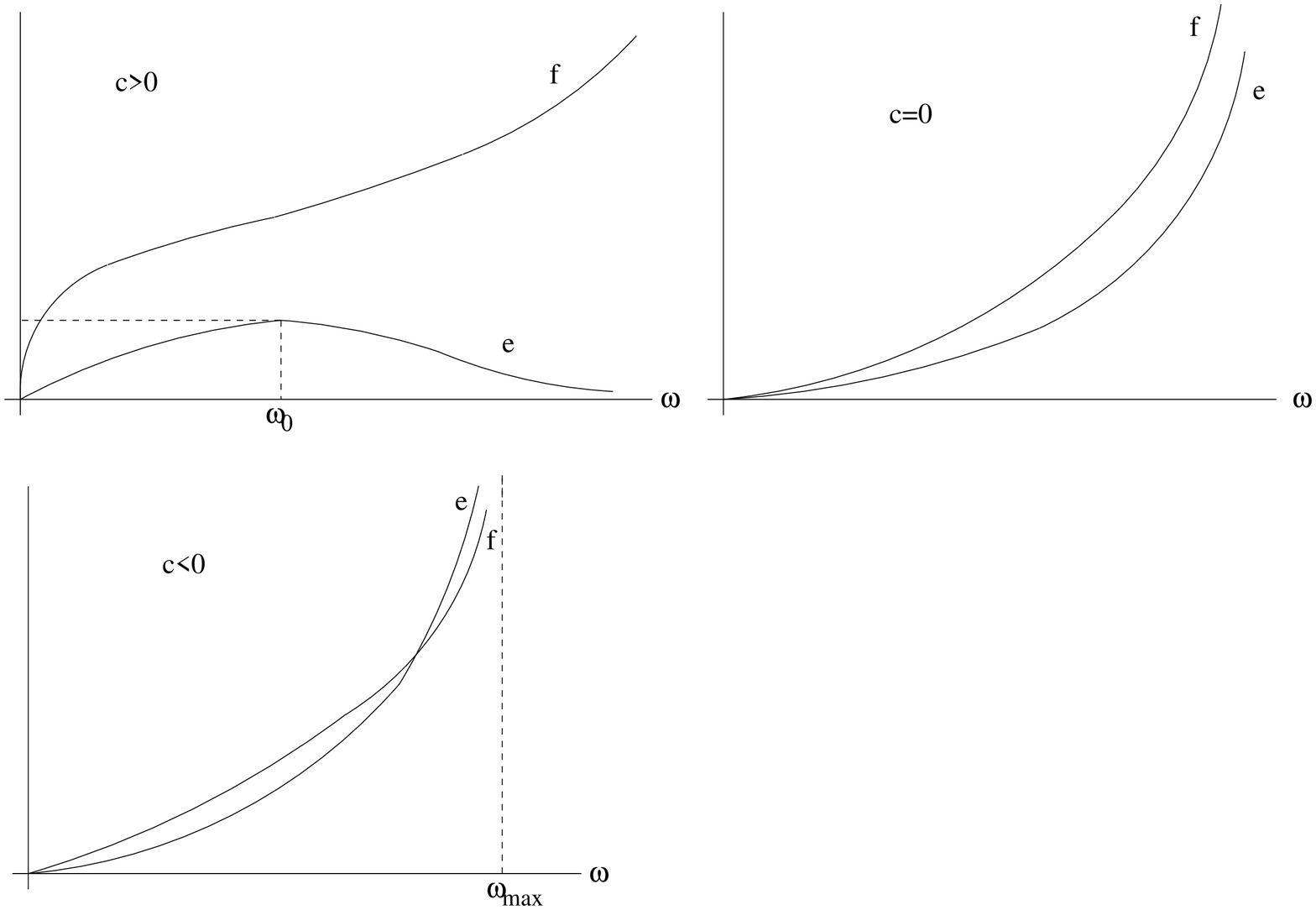}}}
\caption{Qualitative behaviour of $e^{2\protect\sqrt2\phi}$ (curves e) 
and $e^{2\chi}$ (curves f) as functions of $\om$ for type-IIA and
type-IIB.}
\end{center} 
\label{typeII}
\end{figure}

Performing the integration of  
the conformal factor, we find that the metric takes the form
\be
ds^2 ={8 e^{\sqrt{2} \phi}\ov \sinh^2\om \cosh^4\om} d\om^2 
+{e^{-\sqrt{2}\phi}\ov \cosh^2\om} \eta_{\m\n} dx^\m dx^\n \ .
\label{metII}
\ee
As for the 
relation between the variables $r$ and $\om$ in \eqn{metriki}
and \eqn{metII} this turns out to be given by the relation of
differentials
\be
dr = - {2 \sqrt{2} e^{\phi/\sqrt{2}}\ov \sinh\om \cosh^2\om}d\om \ .
\label{chaII}
\ee
Clearly, this integration cannot be performed in closed form and
so we lack the explicit dependence of $\om(r)$. However, the dependence
can easily be spelled out in some limiting cases to which we now turn.

\no\underline{Universal behaviour for small windings}: First
consider the limit of vanishing 
winding field $\om$. Then the asymptotic behaviour of
the fields is 
\be
\om\simeq e^{- (3r/8)^{4/3}}\ ,\qq e^{-2\chi}\simeq e^{-\sqrt{2}\phi}
\simeq \left({3\ov \sqrt{8}}\ r\right)^{2/3}\ ,\qq {\rm as}\quad r\to
\infty\ .
\label{assyII}
\ee
Hence, in the limit of vanishing winding field $\om$, the fields $\phi$
and $\chi$ approach $-\infty$ and the value of the 
constant $c$ plays no r\^ole.
\footnote{The exponential vanishing of 
the winding field $\om$ is related to the fact that setting $\om=0$ 
from the very 
beginning is consistent with the system of eqs. \eqn{syII}. In fact,
then 
the solution \eqn{assyII} for $\chi$ and $\phi$ and  the metric 
in \eqn{metIIlim} below become exact (see also \eqn{exxa} below).}
In addition, the metric becomes
\be
ds^2\simeq dr^2 + \left({3 \ov\sqrt{8}}\ r \right)^{2/3} 
\eta_{\m\n} dx^\m dx^\n\ ,
\qq {\rm as}\quad r\to \infty\ .
\label{metIIlim}
\ee

The behaviour of the solution for larger values of the winding $\om$
does 
depend on the parameter $c$. We distinguish the following three cases:

\no\underline{$c>0$}: In this case 
the reality conditions for the fields $\chi$ and $\phi$ allow for  
$0\leq \om < \infty$.
Then, the string coupling 
$e^{2\chi}$ is a monotonically increasing function of $\om$ in its 
full range of values. 
However, the temperature field squared 
$e^{2\sqrt{2}\phi}$ first increases until 
it reaches a maximum at $\om=\om_0$ and then decreases to zero as 
$\om\to \infty$. 
In the limits of small and large $c$ the constant 
$\om_0$ can be found analytically:
\be
\om_0 = \left\{\begin{array}{lll}
-\frac{1}{4} \ln\left(c\ov 8\right)  & \ \ {\rm for}\  & c \to 0^+\ ,
 \\
\frac{1}{\sqrt{c}}  & \ \ {\rm for}\ & c \to \infty  \ .
\end{array} 
\right.
\label{limi16}
\ee
For intermediate values of $c$, $\om_0$ ranges between the above two
extremes.
The maximum value of $e^{2\sqrt{2}\phi}$ is given by 
\be
e^{2\sqrt{2}\phi}\Big|_{\om=\om_0}= \ \ha \sinh^2\om_0\ .
\ee
The asymptotic behaviour of the various fields is
found by noting that due to \eqn{chaII} we have the relation
\be 
r \simeq {32\cdot 2^{3/4}\ov 7 c^{1/4}} e^{-7\om/2}\ ,
\qq {\rm as} \quad r\to 0^+
\quad {\rm and}\quad \om \to \infty\ ,
\label{liiII}
\ee
up to an additive constant that has been fixed in an obvious way.
Then, 
\be
e^{-\sqrt{2}\phi} \simeq \left({16 c^{3/2}\ov 7 r}\right)^{2/7} \ ,\qq
e^{-2\chi} \simeq \left({7\ov \sqrt{2}} c^2 r\right)^{2/7}\ ,\qq {\rm
as}
\quad r\to 0^+\ .
\ee
Similarly, the asymptotic expression for the metric is 
\be
ds^2\simeq dr^2 + \left({7\ov \sqrt{2}} c^2 r\right)^{2/7} \eta_{\m\n}
dx^\m dx^\n\ ,\qq {\rm as} \quad r\to 0^+\ .
\ee

\no\underline{$c=0$}: As before,
the reality conditions for the fields $\chi$ and $\phi$ allow for 
$0\leq \om < \infty$. However, both fields  
$e^{2\chi}$ and  $e^{2\sqrt{2}\phi}$ are now monotonically increasing
functions
of $\om$ in the entire range of values.
The asymptotic behaviour of the various fields is
found by noting that due to \eqn{chaII} we have the relation
\be 
 r \simeq {32 \ov 5} e^{-5\om/2}\ ,\qq {\rm as} \quad r\to 0^+
\quad {\rm and}\quad \om \to \infty\ .
\label{liiII2}
\ee
Then, 
\be
e^{-\sqrt{2}\phi} \simeq \ha (5 r)^{2/5}\ ,\qq
e^{-2\chi} \simeq {1\ov 8} (5r)^{6/5}\ ,\qq {\rm as} \quad r\to 0^+\ .
\ee
Similarly, the asymptotic expression for the metric is 
\be
ds^2\simeq dr^2 + {1\ov 8} (5r)^{6/5} \eta_{\m\n}
dx^\m dx^\n\ ,\qq {\rm as} \quad r\to 0^+\ .
\ee

\no\underline{$c<0$}: In this case
the reality conditions for the fields $\chi$ and $\phi$ do not allow to 
exceed a maximum value for the winding field, i.e., 
$0\leq \om \leq \om_{\rm max}$.
In this range, $e^{2\chi}$ and $e^{2\sqrt{2}\phi}$ are monotonically 
increasing functions of $\om$.
Similarly, in the limit of small or large $|c|$ the
constant $\om_{\rm max}$ can be found analytically, as before, 
\be
\om_{\rm max} = \left\{\begin{array}{lll}
-\frac{1}{4} \ln\left(-c\ov 8\right) & \ \ {\rm for}\  & c \to 0^-\ ,
 \\
e^{c-1\ov 2} & \ \ {\rm for}\  & c \to - \infty  \ ,
\end{array} 
\right.
\label{lima6}
\ee
whereas for intermediate values of $c$, $\om_{\rm max}$ 
ranges between the two above extremes.
The asymptotic behaviour of the various fields can be  
found by noting that due to \eqn{chaII} we have the relation
\be 
 r \simeq {8\ov 3 {\rm s_m^{3/4}} {\rm c_m^{7/4}} }
(\om_{\rm max} -\om)^{3/4}\ ,\qq {\rm as} \quad
r\to 0^+ \quad {\rm and}\quad \om \to \om_{\rm max}^- \ ,
\label{l1II}
\ee
by introducing the notation 
\be
\label{noott}
{\rm s_m}= \sinh \om_{\rm max} \ , \qq  {\rm c_m}= \cosh \om_{\rm max}\
. 
\ee
Then, 
\be
e^{-\sqrt{2}\phi}\simeq \left({3 {\rm c_m}\ov 2\sqrt{2}}\ r\right)^{2/3}
\ ,\qq
e^{-2\chi} \simeq 
\left({3\ov 2\sqrt{2} {\rm c^2_m}}\ r\right)^{2/3} 
\ ,\qq {\rm as} \quad r\to 0^+\ .
\ee
Similarly, the asymptotic expression for the metric is 
\be
ds^2\simeq dr^2 + 
\left({3\ov 2\sqrt{2} {\rm c^2_m}}\ r\right)^{2/3} 
\eta_{\m\n} dx^\m dx^\n\ ,\qq {\rm as} \quad r\to 0^+\ .
\ee

In all cases above, we note that the main qualitative difference 
is in the shape of 
the function $e^{2\sqrt{2}\phi}$ representing the temperature field 
square: for $c\leq 0$
it starts from zero and evolves 
monotonically to $+\infty$ following the range of $r$ 
from $\infty$ to 0, 
whereas for $c>0$ we see a dramatic change in that 
the function starts and ends
at zero, thus developing a maximum along the way. 
The end point at $r=0$ corresponds to a curvature singularity. 
As we will see in section 6, there are physical ways 
to differentiate among
the allowed values of $c$ by requiring consistent propagation of 
a quantum test particle on the corresponding supergravity 
backgrounds which have a curvature singularity.
Of course, string theory (and not supergravity alone)
holds the ultimate answer for the physical relevance of our solutions 
in thermodynamics.

\subsection{A hybrid type-II sector}
There is another truncation of the domain wall equations with  
$H_1=0$, as in type-II, which is exactly solvable. 
We set for this purpose 
$H_2 = \pm H_3$ and observe that the full system of equations is
consistent provided that $\phi_2 = \phi_3$. Actually, this sector
can be viewed as a hybrid of type-IIB and IIA in that we impose 
the ``diagonal" constraint $H_2 = \pm H_3$ instead of choosing 
the axes $H_2=0$
or $H_3 =0$ respectively in $H$-field space; as such, it is a hybrid
truncation. This truncation amounts to choosing 
the self-dual radius so that there
is no dinstinction between the type-IIA and type-IIB theories.

We proceed further by setting
\be
\phi_1 = \chi ~, ~~~~ \phi_2 = \phi_3 = {1 \over \sqrt{2}}\phi ~, ~~~~ 
H_2 = \pm H_3 = {1 \over 2\sqrt{2}} {\rm sinh}\omega ~ 
\ee
and so $z_2 = \pm z_3 \equiv z$ can be chosen as 
\be
z = {1 \over \sqrt{2}} {\rm tanh}{\omega \over 2} \ ,
\ee
in order for the kinetic terms of the fields $\chi$, $\phi$ and 
$\omega$ to assume their canonical form. 
As before, the temperature field is $T\sim e^{\sqrt{2} \phi}$.
Then, the prepotential \eqn{supereal} and the potential \eqn{dinamiko}
become 
\be
\wt_{{\rm hyb}} = {1 \over 2}\left(e^{\chi +\sqrt{2} \phi} 
+{1 \over 2} e^{-\chi} 
{\rm sinh}^2 \omega \right)
\ee
and 
\be
V_{{\rm hyb}} = {1\ov 32} \sinh^2\om \left(e^{-2 \chi} 
(1+\cosh^2\om)-8 e^{\sqrt{2}\phi} \right)\ ,
\ee
respectively. We see that $\om$ will become tachyonic, if at $\om=0$ we
have 
$e^{-\sqrt{2}\phi -2 \chi}<4$.

The truncated system of equations is 
\ba
{d \chi \over dr} & = & -{1 \over 2 \sqrt{2}} \left(e^{\chi +\sqrt{2}
\phi} 
- {1 \over 2} e^{-\chi} {\rm sinh}^2 \omega \right)\ , 
\nonumber\\
{d \phi \over dr} & = & -{1 \over 2} e^{\chi +\sqrt{2} \phi} \ ,
\label{syhybII}\\ 
{d \omega \over dr} & = & -{1 \over 4\sqrt{2}}e^{-\chi} \sinh 2\omega  ~
.
\nonumber 
\ea
The equation for the 
conformal factor of the metric \eqn{metriki} can be easily integrated;
it gives 
$A=-\chi-\ln (\cosh \om)$, up to a constant that can be absorbed into
a redefinition of the $x^\m$'s. Notice that 
the result is not the same as in the 
genuine type-II case that we examined before.
The system of equations \eqn{syhybII} is 
invariant under $\om\to -\om$, and as before we only need to
consider the case of
$\om\ge 0$. 
Note also that, unlike the genuine type-II case, 
the field $\chi$ cannot be viewed  
as a Liouville field in this hybrid sector that   
provides the ``gravitational dressing" of a simpler massive model
for the scalar fields $\phi$ and $\omega$. 

Luckily, the general solution can be found in parametric form by
employing $\omega$ as an independent variable, as before. We find 
the following family of BPS solutions 
\ba
e^{-2\sqrt{2}\phi} & = & b + 
2a \left({\rm \ln}\left({\rm coth}{\omega \over 2}
\right) - {1 \over {\rm cosh} \omega}\right) \ ,
\nonumber\\
e^{2 \chi} & =&  {a \over 2 {\rm cosh}\omega} e^{\sqrt{2} \phi} \ ,
\ea
where $a$ and $b$ are integration constants with $a$ being positive for 
reality reasons. For convenience, in the rest of this subsection, we set 
the positive constant $a=2$, since its precise
value can be adjusted by field redefinitions and hence it has no 
physical relevance. The other integration constant $b$ cannot be
fixed mathematically, but we will see later that physical considerations
in the context of supergravity may impose some restrictions on it. Of
course,
as before, it is string theory that holds the answer for having an
acceptable geometrical background  for domain walls. 
The behaviour of the functions $e^{2\sqrt{2}\phi}$ and $e^{2 \chi}$ for
the 
three choices of $b$ is sketched in Figure~2.
Again, as $\om\to 0$, the temperature field behaves as 
$T^{-2}\sim e^{-2\sqrt{2}\phi}\simeq 2 \ln \om^{-2}\to \infty$ 
and $e^{-2\chi}\sim e^{-\sqrt{2}\phi}\to\infty$ so that $\om$ 
cannot be tachyonic.

\begin{figure}[ht]
\begin{center}
{\scalebox{.6}{\includegraphics{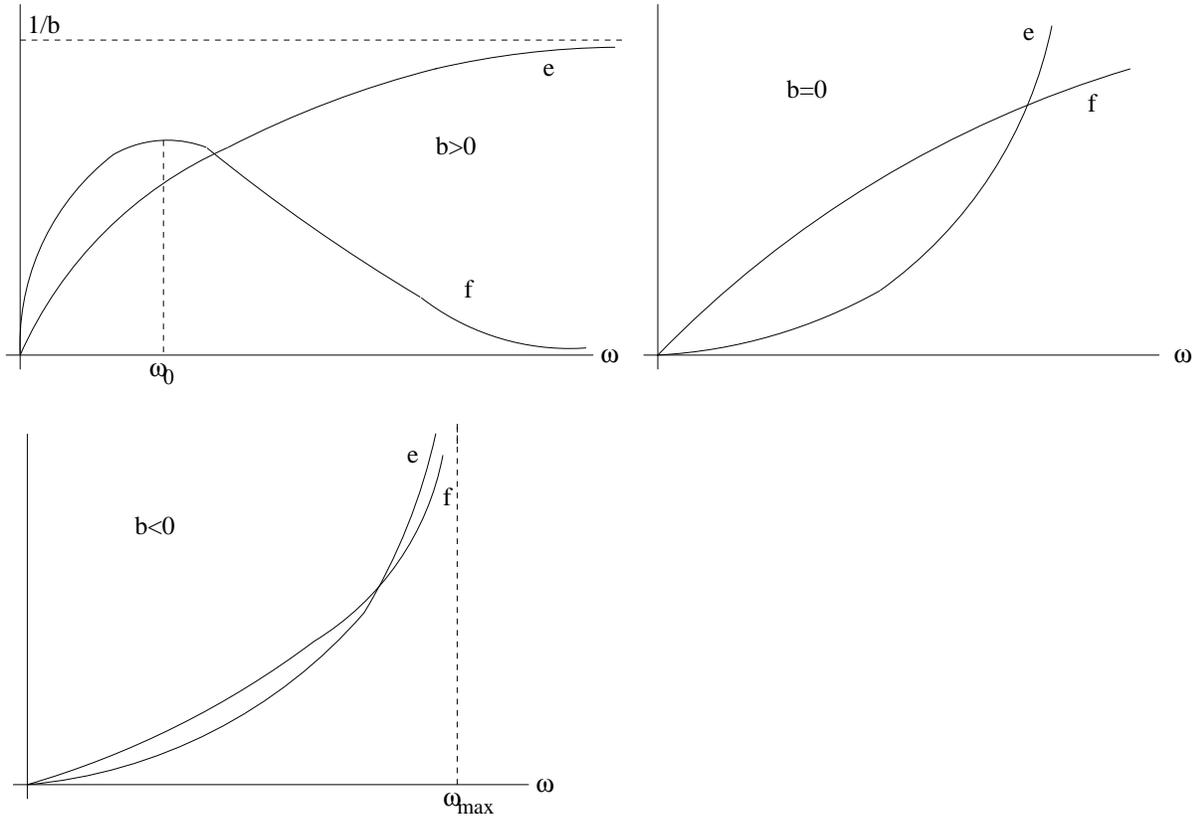}}}
\caption{Qualitative behaviour of $e^{2\protect\sqrt{2}\phi}$ (curves e) 
and $e^{2\chi}$ (curves f) as functions of $\om$ for the hybrid
type-II.}
\end{center} 
\label{hybridII}
\end{figure}

After computing the conformal factor we may write the metric as
\be
ds^2 = {4a e^{\sqrt{2}\phi}\ov \sinh^2\om \cosh^3\om}d\om^2 +
{2 e^{-\sqrt{2} \phi}\ov a \cosh\om} \eta_{\m\n} dx^\m dx^\n \ .
\label{metHyb}
\ee
The connection between the variables $r$ and $\om$ in \eqn{metriki}
and \eqn{metHyb} is given by the relation of differentials
\be
dr = - {2 \sqrt{a} e^{\phi/\sqrt{2}}\ov \sinh\om \cosh^{3/2}\om}d\om \ .
\label{chahybII}
\ee
As before, the integration cannot be performed in closed form, apart
from
a few limiting 
cases, as we will see next.

\no\underline{Universal behaviour for small windings}:  
Let us consider first the case of vanishing
winding field $\om$. 
It turns out that the asymptotic behaviour of
the fields is 
\be
\om\simeq 2 e^{- (3r/8)^{4/3}}\ ,\qq e^{-2\chi}\simeq e^{-\sqrt{2}\phi}
\simeq \left({3\ov \sqrt{8}}\ r\right)^{2/3}\ ,\qq {\rm as}\quad r\to
\infty\ .
\label{hybII}
\ee
Hence, in the limit of vanishing winding field $\om$, the fields $\phi$
and $\chi$ approach $-\infty$ and the value of the 
constant $b$ plays no r\^ole.\footnote{A similar comment, as in 
footnote 6, applies here as well.}
The metric becomes
\be
ds^2\simeq dr^2 + \left({3 \ov\sqrt{8}}\ r \right)^{2/3} 
\eta_{\m\n} dx^\m dx^\n\ ,
\qq {\rm as}\quad r\to \infty\ .
\label{methybII}
\ee
Note that this asymptotic behaviour is almost identical to the one found 
for the pure type-II case. A similar behaviour will also exist in 
the heterotic case to be discussed in the next section. There we will
see 
that it is related to an exact solution with zero winding present in 
all cases.

The behaviour of the solution for larger values of the winding $\om$,
however, depends on the parameter $b$. As in the genuine type-II
solution \eqn{chiII}, we also distinguish here the following three
cases:

\no\underline{$b>0$}: In this case 
the reality conditions for the fields $\chi$ and $\phi$ allow for 
$0\leq \om < \infty$.
Then, the temperature field square 
$e^{2\sqrt{2} \phi}$ is a monotonically increasing function of $\om$
until
it reaches the asymptotic value $1/b$.
However, the string coupling $e^{2\chi}$ first increases until 
it reaches a maximum at $\om=\om_0$, and then decreases to zero as 
$\om\to \infty$. 
In the limits of small and large $b$, the constant 
$\om_0$ can be found analytically
\be
\om_0 = \left\{\begin{array}{lll}
-\frac{1}{3} \ln\left(3b\ov 16\right) & \ \ {\rm for}\  & b \to 0^+\ ,
 \\
\sqrt{2\ov b} & \ \ {\rm for}\  & b \to \infty  \ ,
\end{array} 
\right.
\label{lihyb6}
\ee
whereas for intermediate values of $b$, $\om_0$ 
ranges between the above two extremes.
The maximum value of $e^{2\chi}$ is given by 
\be
e^{2\chi}\Big|_{\om=\om_0}= \ {\sinh\om_0 \ov \sqrt{2 \cosh\om_0}} \ .
\ee
The asymptotic behaviour of the various fields is
extracted by noting that, due to \eqn{chahybII}, we have the relation
\be 
r \simeq {32\ov 5 b^{1/4}} e^{-5\om/2}\ ,
\qq {\rm as} \quad r\to 0^+
\quad {\rm and}\quad \om \to \infty\ .
\label{lihhI}
\ee
Then, we find that
\be
e^{-2\sqrt{2}\phi} \simeq b + {1\ov 6} (5 b^{1/4} r)^{6/5} \ ,\qq
e^{2\chi} \simeq \ha \left(5 r\ov  b\right)^{2/5}\ ,\qq {\rm as}
\quad r\to 0^+\ .
\ee
Similarly, the asymptotic expression for the metric is 
\be
ds^2\simeq dr^2 + \ha ({5 b^{3/2}} r )^{2/5} \eta_{\m\n}
dx^\m dx^\n\ ,\qq {\rm as} \quad r\to 0^+\ .
\ee

\no\underline{$b=0$}: As before,
the reality conditions for the fields $\chi$ and $\phi$ allow for 
$0\leq \om < \infty$. However, both fields  
$e^{2\chi}$ and  $e^{2\sqrt{2}\phi}$ are now monotonically increasing
functions
of $\om$ in the entire range of values.
The asymptotic behaviour of the various fields is
found using 
\be 
r \simeq {32 \ov 7} \left({3 \over 2} \right)^{1/4} 
e^{-7\om/4}\ ,\qq {\rm as} \quad r\to 0^+
\quad {\rm and}\quad \om \to \infty\ ,
\label{lihyb2}
\ee
which is due to \eqn{chahybII}. Then, we find that
\be
e^{-2 \sqrt{2}\phi} \simeq {7\ov 24} \left({7^5 \over 54}\right)^{1/7} 
r^{12/7}\ ,\qq
e^{-2\chi} \simeq \left({98 \over 81} \right)^{1/7} 
r^{2/7}\ ,\qq {\rm as} \quad r\to 0^+\ .
\ee
Similarly, the asymptotic expression for the metric is 
\be
ds^2\simeq dr^2 + {7\ov 24} \left(3\cdot 7^3\ov 4\right)^{1/7}
r^{10/7} \eta_{\m\n} dx^\m dx^\n\ ,\qq {\rm as} \quad r\to 0^+\ .
\ee

\no\underline{$b<0$}: In this case
the reality conditions for the fields $\chi$ and $\phi$ do not allow to 
exceed a maximum value for the winding field, i.e., 
$0\leq \om \leq \om_{\rm max}$.
In this range, $e^{2\chi}$ and $e^{2\sqrt{2}\phi}$ are monotonically 
increasing functions of $\om$.
In the limit of small and large 
$|b|$ we find for $\om_{\rm max}$ the analytic 
expressions
\be
\om_{\rm max} = \left\{\begin{array}{lll}
-\frac{1}{3} \ln\left(-3b\ov 32\right) & \ \ {\rm for}\  & b \to 0^-\ ,
 \\
2 e^{b/4-1} & \ \ {\rm for}\  & b \to - \infty  \ .
\end{array} 
\right.
\label{lhyyba6}
\ee
For intermediate values of $b$, $\om_{\rm max}$ ranges between the 
two above extremes.
The asymptotic behaviour of the various fields is
extracted by noting that, due to \eqn{chahybII}, we have the relation
\be 
 r \simeq {8\ov 3 {\rm s^{3/4}_m} {\rm c_m} }
(\om_{\rm max} -\om)^{3/4}\ ,\qq {\rm as} \quad
r\to 0^+ \quad {\rm and}\quad \om \to \om_{\rm max}^- \ ,
\label{lyybI}
\ee
where we have used the same notation as in \eqn{noott}. Then, 
\be
e^{-\sqrt{2}\phi}\simeq \left({3 \ov 2\sqrt{2 {\rm c_m}}}\ 
r\right)^{2/3} \ ,\qq
e^{-2\chi} \simeq 
\left({3  {\rm c_m} \ov 2\sqrt{2}}\ r\right)^{2/3} 
\ ,\qq {\rm as} \quad r\to 0^+\ .
\ee
Similarly, the asymptotic expression for the metric is 
\be
ds^2\simeq dr^2 + 
\left({3 \ov 2\sqrt{2} {\rm c_m^2}}\ r\right)^{2/3} 
\eta_{\m\n} dx^\m dx^\n\ ,\qq {\rm as} \quad r\to 0^+\ .
\ee

Note that in the hybrid type-II sector, as for the genuine type-II
solution, 
the end point at $r=0$ represents a curvature singularity.
We will see in section 6 that the three different cases, 
which correspond to the values of $b$ $>0$, 0 or $<0$,
can be distinguished using certain physical criteria.

Finally, note that for all type-II sectors (IIA, IIB and the hybrid
type-II), 
$\om$ decreases monotonically from its value at $r=0$ (which may be 
infinite) to $\om=0$ at $r=\infty$. In particular, this implies 
that $\om(r)$  cannot be a periodic function of $r$, and hence $r$ 
necessarily is a non-compact coordinate.

\section{Heterotic sector} 
\setcounter{equation}{0}
 
The heterotic limit is obtained by setting $H_2 = H_3 = 0$, while 
keeping $H_1 \equiv H$ free to vary. 
It then follows from \eqn{fi2} and \eqn{fi3} that $\phi_2=\phi_3$ up
to an irrelevant additive constant that we will ignore.
Trading the 
$z$ corresponding to $H$ for a winding field 
$\omega$ and introducing appropriate 
variables 
\be
\phi_1 = \chi ~, ~~~~ \phi_2 = \phi_3 = {\phi \ov \sqrt{2}} ~, ~~~~ 
z = {\rm tanh} {\omega \over 2} ~, 
\ee
for which $2H = {\rm sinh} \omega$, 
we obtain the expression for the truncated prepotential 
\be
\wt_{{\rm het}} = {1 \over 2} e^{\chi} \left( e^{\sqrt{2}\phi} - 
{\rm sinh}(\sqrt{2} \phi) {\rm sinh}^2 \omega \right) , 
\ee
and the potential 
\be
V_{{\rm het}} = {1 \over 8} e^{2 \chi} \sinh^2\om \left(
{\rm cosh}(2\sqrt{2} \phi)\cosh^2 \omega  - 3\right) \ .  
\ee
We note for completeness that $e^\chi$ is the string coupling and
the temperature field is  again $e^{\sqrt{2}\phi}$ 
(normalized in heterotic string units that remove the $\chi$-dependence)
and
the kinetic terms of the fields
$\chi$, $\phi$ and $\omega$ assume their canonical form with 
our choice of normalizations. We see that $\om$ can be tachyonic only if 
$\cosh (2\sqrt{2}\phi) < 3$ at $\om=0$.

We remark that although the potential $V_{\rm het}$ is invariant 
under $\om\to -\om$ and $\phi\to -\phi$, the prepotential 
$\wt_{\rm het}$ is invariant under the first transformation only. 
The second transformation corresponds to T-duality in the temperature 
radius. Since the prepotential breaks this temperature T-duality one 
expects the same to be true for non-trivial solutions of our equations.

The truncated system of differential equations is 
\ba
{d \chi \over dr} & = & -{1 \over 2\sqrt{2}} e^{\chi} \left(
e^{\sqrt{2} \phi} - {\rm sinh}(\sqrt{2} \phi) {\rm sinh}^2 \omega 
\right) , \nonumber \\
{d \phi \over dr} & = & -{1 \over 2} e^{\chi} \left(
e^{\sqrt{2} \phi} - {\rm cosh}(\sqrt{2} \phi) {\rm sinh}^2 \omega 
\right) , 
\label{het3}\\
{d \omega \over dr} & = & {1 \over 2\sqrt{2}} e^{\chi} 
{\rm sinh}(\sqrt{2} \phi) \sinh 2 \om ~. 
\nonumber 
\ea
The equation for the 
conformal factor of the metric \eqn{metriki} is easily integrated and
gives 
$A=-\chi$, as in the case of the genuine type-IIA or type-IIB theories.
Also, similarly to this case, we may drop the $\chi$-dependence
from the right hand side of the resulting first order system by 
simply changing variables to $\zeta$ defined by 
$d\zeta = e^\chi dr$. Again, $\chi$ may be viewed
as a Liouville field whose coupling provides the ``gravitational 
dressing" of an auxiliary massive model of 
the fields $\phi$ and $\omega$ with
Lagrangian density
\be
{\cal L}^{\prime}(\phi, \omega) = -{1 \over 2} (\partial \phi)^2 
- {1 \over 2} (\partial \omega)^2 - {1 \over 8}  \sinh^2\om \left(
{\rm cosh}(2\sqrt{2} \phi)\cosh^2 \omega  - 3\right) \ .  
\ee
This prescription produces naturally the canonical kinetic term for
the field $\chi$ and dresses the potential into $V_{{\rm het}}$.   

It turns out that it is not possible to solve the domain wall equations
in closed form in the heterotic sector. 
Nevertheless, we can still determine the leading small 
$\omega$ asymptotics, similarly to the analogous universal 
behaviour for the type-II theories. 
This corresponds to large $r$. As it 
turns out, $\omega$ is again exponentially suppressed 
as $r\to\infty$, while $e^\chi$ and $e^{\sqrt{2}\phi}$ 
have a power-law behaviour. Hence, we can safely drop the 
$\sinh^2\omega$ part in the first two equations in \eqn{het3} and 
replace $\sinh 2\omega$ by $2\omega$ in the third equation. 
The resulting system is easily integrated and yields in this limit 
\be
 e^{-\sqrt{2}\phi} \simeq 2 C r^{2/3} \ , \quad 
e^{-2\chi} \simeq {9\over 32 C^2} r^{2/3} \ , \quad 
\omega \simeq \omega_0\ e^{ - C^2 r^{4/3}}\ , \qq
\quad {\rm as}\ r\to\infty \ ,
\ee
where $C$ and $\om_0$ are two positive constants of integration.
Note that $\cosh (2\sqrt{2}\phi) \simeq 2 C^2 r^{4/3} 
\simeq 2 \ln \om^{-2} \to \infty$ as $\om\to 0$ and $\om$ cannot be
tachyonic.
As one decreases the value of the constant $\omega_0$, the range of 
approximate validity of this asymptotic solution extends 
to larger intervals of $r$. Eventually, if $\omega_0=0$ 
one obtains a particular exact solution\footnote{We note that this is an
exact solution for the 
type-II sector as well. Indeed, when $\om=0$, the three different 
truncated systems of equations \eqn{syII}, \eqn{syhybII}
and \eqn{het3} become identical. Note that for the corresponding 
type-II solution we had chosen the integration constant $C=3^{2/3}/4$.}
\be
 e^{-\sqrt{2}\phi} =2 C r^{2/3} \ , \qq
e^{-2\chi} = {9\over 32 C^2} r^{2/3} \ , \qq
\omega =0 \ , 
\label{exxa}
\ee
which is valid for all values of $r$. 
However, since we lack the most general 
solution, even in some parametric form $\phi(\omega)$, we will only 
provide results about the dominant behaviour of the fields in the
weak coupling (i.e., $\chi \rightarrow -\infty$) and the strong coupling 
(i.e., $\chi \rightarrow + \infty$) regions. In fact, we will 
succeed to identify
these regions and perform an in depth analysis in their vicinity using 
standard techniques from dynamical systems. The remaining 
part of the two-dimensional space $(\phi, \omega)$ can only 
be studied numerically.

Before delving into details, we study a bit more the exact solution for 
$\om=0$. This solution represents a straight orbit in the 
parameter space $(\phi, \omega)$
with $\omega=0$ everywhere, extending from $\phi=-\infty$
(identified as a weak coupling point) to $\phi=+\infty$
(identified as a strong coupling point), 
as $r$ ranges from  
$+\infty$ to 0. The Liouville coordinate $\zeta$, which 
trades $r$ as $d\zeta = e^\chi dr$, is given for this special
solution by 
\be
{\zeta} = 2 \sqrt{2} C r^{2/3} ~, 
\ee  
where we absorbed an integration constant into $\zeta$.

Introducing small fluctuations $\delta \phi$ and $\delta \omega$ 
around this special solution, to test
its stability, we find to first order that they satisfy the linearized
system of equations  
\ba
{d \over d\zeta}\delta \phi & = & -{1 \over \zeta} \delta \phi\ , 
\nonumber\\
{d \over d\zeta} \delta \omega & = & \left({1 \over 2\zeta} - 
{\zeta \over 4}\right) \delta \omega \ ,
\ea
which are written for convenience using $\zeta$. Therefore, 
we have the following solution of the linearized problem
\be
\delta \phi = {A \over \zeta} ~, ~~~~~ \delta \omega = B \sqrt{\zeta}
e^{-{\zeta}^2 /8} ~, 
\ee
where $A$, $B$ are some other constants. These fluctuations are 
small provided that $\zeta \rightarrow \infty$, otherwise the linearized
approximation is not valid. We conclude this analysis by noting that
the exact solution we found is stable against perturbations in the 
parameter space $(\phi, \omega)$ provided that we are in the weak
coupling region where $\phi \rightarrow - \infty$. In fact, as we
will see next, the point $(-\infty, 0)$ is an asymptotic 
critical point in the two-dimensional parameter space, and so we
can talk about asymptotic stability in its vicinity.

\no\underline{The critical points}: It is more convenient to work with
the coordinate $\zeta$ instead of $r$ in order to drop the 
$\chi$-dependence from the right hand side of the equations. Then, 
as in the theory of dynamical systems, we have to study the solutions
of the system
\be
{d\phi \over d\zeta} = P(\phi, \omega) ~, \qq
{d\omega \over d\zeta} = Q(\phi, \omega)\ ,
\ee
either as functions of $\zeta$, i.e., as $\phi(\zeta)$ and
$\omega(\zeta)$, 
or in a parametric form $ \phi(\omega)$. For the moment we have not
prescribed a set of physical boundary conditions that select certain
orbits among the infinite many that arise in the two-dimensional
parameter space $(\phi, \omega)$, as this will be done in the next 
section. Here, we only focus on general properties of the solutions
around the critical points of the system.

Recall that the critical points are defined as the common zeros of
the functions $P$ and $Q$. These occur in the present case either
when $\phi = 0$ and ${\rm sinh}\omega = \pm 1$ or when 
$e^{\sqrt{2}\phi} = 0$ and $\omega = 0$. The second case 
has already been analyzed and lies in the weak coupling region of 
the exact solution that was described above with $\phi$ approaching 
asymptotically $-\infty$. Therefore, we focus on
the other critical points 
\be
\phi_0 = 0 ~, ~~~~~ \omega_0^{\pm} = {\rm \ln}(\sqrt{2} \pm 1) ~, 
\label{crii}
\ee
which appear symmetrically on the 
$\omega$-axis because of the invariance of the equations 
under $\omega \rightarrow -\omega$. 
Then, the dilaton equation in \eqn{het3} is easily integrated to give 
at these critical points
\be
\chi 
= -{1\ov 2 \sqrt{2}} \zeta  = \ln\left(2 \sqrt{2}\ov r\right)\ ,
\label{liined}
\ee
where we have absorbed integration constants
into $\zeta$ and $r$ in an obvious way.
Hence, the dilaton 
field exhibits a logarithmic dependence on $r$ and a linear in $\zeta$.
Recall now that the string frame metric in four dimensions 
is obtained by \eqn{metriki} after 
multiplying with $e^{2 \chi}$. Hence, the resulting backgound has a 
flat string metric and a linear dilaton, which is the exact solution 
preserving half of the supersymmetries found earlier in \cite{ak, adk}.
Note that, in this case, it makes sense to pass from the Einstein 
to the string frame since,
when $\phi$ and $\om$ assume their critical values \eqn{crii},
we are left with only massless fields (graviton and dilaton) that 
couple to a 2-dimensional string world-sheet action.

Next, if we linearize around these 
critical points \eqn{crii} by introducing small fluctuations $\delta
\phi$ and
$\delta \omega$, we will find the following evolution for them 
\be
{d \over d \zeta} 
\left(\begin{array}{c} 
\delta \phi \\
            \\
\delta \omega \end{array}\right) 
= -{1 \over \sqrt{2}}  
\left(\begin{array}{ccc}
1 & & \mp 2\\
  & &      \\
\mp 2 & & 0 \end{array}\right) 
\left(\begin{array}{c} 
\delta \phi \\
            \\
\delta \omega \end{array}\right)\ ,
\ee
whereas for the dilaton ${d(\d\chi)/ d\zeta}=0$.
Computing the two eigenvalues of the corresponding matrix we have 
in either case of ${\omega}_0^{\pm}$ that  
\be
\lambda_{\pm} = {1 \over 2\sqrt{2}} (-1 \pm \sqrt{17}) ~,  
\ee
and since $\lambda_+ >0$ and $\lambda_- <0$, we see that the pair
of critical points $(0, \omega_0^{\pm})$ are both saddle points. 
We may solve the linearized system of equations and obtain 
\ba
\delta \phi & =& A e^{\lambda_+ \zeta} + B e^{\lambda_- \zeta} ~, 
\nonumber\\
\delta \omega &=& \pm \sqrt{2} \left({A \over \lambda_+} 
e^{\lambda_+ \zeta} + {B \over \lambda_-} e^{\lambda_- \zeta} 
\right)  \ ,
\ea
with $\pm$ depending on the choice of critical point 
$\omega_0^{\pm}$, whereas the dilaton does not change, to linear order,
from its value in \eqn{liined}.

If we demand that
the critical points are reached as 
$\zeta \rightarrow +\infty$, we conclude
for the integration constants that $A=0$.
Hence, the variations of $\phi$ and $\om$ 
have a power law decay, namely
\be
\delta \phi = \mp { \sqrt{17} +1 
\over 4} \delta \omega = B \left(2 \sqrt{2}\ov r\right)^{\sqrt{17}+1}\ ,
\qq {\rm as}\quad r \rightarrow \infty\ .
\label{ght1}
\ee
On the other hand, if we demand instead, that 
the critical points are reached as 
$\zeta \rightarrow -\infty$, we conclude
for the integration constants that $B=0$.
Now eq. \eqn{ght1} is replaced by 
\be
\delta \phi = \pm {\sqrt{17}-1 
\over 4} \delta \omega = A \left(r\ov 2 \sqrt{2}\right)^{\sqrt{17}-1}\ ,
\qq {\rm as}\quad r \rightarrow 0^+\ .
\label{ght2}
\ee
Note the different behaviour of the 
two solutions around the critical points \eqn{crii} as given by 
\eqn{ght1} and \eqn{ght2}. In particular, from \eqn{liined},
they correspond to weak and 
strong string couplings, respectively. 
We will see that if trajectories in the entire
$\phi$--$\om$ plane are considered, both solutions will appear.

\no\underline{Strong coupling points}: Having extracted the behaviour
of the fields around the critical points, which can be either at weak or 
at strong coupling, we will examine next their behaviour in the 
vicinity of the other strong coupling points, where
the dilaton field $\chi$ tends to $+\infty$. We have identified three 
such regions in the parameter space $(\phi, \omega)$, which of course
have a mirror counterpart under $\omega \rightarrow -\omega$. 
They are (cf. figure 3 below):
\ba
{\rm region ~1}&:& ~~~~~~ \phi \rightarrow + \infty ~~ {\rm and} ~~ 
\omega \rightarrow 0\ , \\
{\rm region ~2}&:& ~~~~~~ \phi \rightarrow + \infty ~~ {\rm and} ~~ 
\omega \rightarrow \pm \infty \ ,\\
{\rm region ~3}&:& ~~~~~~ \phi \rightarrow - \infty ~~ {\rm and} ~~ 
\omega \rightarrow \pm \infty \ .
\ea
To justify this claim 
and extract the details of the fields in these regions, we shall treat
each case separately.  

First, it is convenient to examine the heterotic equations in the
limit $\phi \rightarrow +\infty$, which is common to both regions 1 and
2.
Using the coordinate $\zeta$ (instead of $r$) we obtain in this limit
the simplified system
\ba
{d\phi \over d\zeta} &=& \sqrt{2} 
{d\chi \over d\zeta} \ = \ - {1 \over 2} e^{\sqrt{2} \phi} 
\left(1 - {1 \over 2} {\rm sinh}^2 \omega \right) , \nonumber
\\
{d\omega \over d\zeta} & =& {1 \over 2\sqrt{2}} e^{\sqrt{2}\phi}
{\rm sinh}\omega {\rm cosh}\omega  
\ea
and so we see immediately that $\sqrt{2} \chi = \phi$ (up to an additive
constant).
The remaining equations can be easily integrated to yield the 
expressions
\ba
e^{-\sqrt{2} \phi} & = & {3 \over 2a\sqrt{2}} (b-a \zeta )^{1/3}
\left(1 - (b-a\zeta )^{2/3}\right) , \nonumber
\\
{\rm cosh}^2 \omega &=& {1 \over (b-a\zeta )^{2/3}} ~,  
\ea
in terms of some constants $a>0$ and $b$. 
However, recall at this point that we are considering the behaviour
of the equations for $\phi \rightarrow +\infty$, and so we should
expand ${\rm exp}(-\sqrt{2} \phi)$ around zero. Naturally, we face 
two possibilities, which will be identified with regions 1 and 2.

Region 1 follows in this context by assuming $(b-a\zeta )^{2/3} 
\simeq 1$. Defining the parameter $\zeta_1 = (b-1)/a$ we have
\be
e^{-\sqrt{2}\phi} \simeq {1 \over \sqrt{2}} (\zeta -\zeta_1) ~, ~~~~ 
\omega \simeq \pm \sqrt{{2a \over 3}} (\zeta - \zeta_1)^{1/2} ~; ~~~~ 
{\rm as} ~~ \zeta \rightarrow \zeta_1^+  
\ee
and so $\omega \rightarrow 0$ as advertized for region 1. Since  
$\sqrt{2} \chi =  \phi$ (up to a constant shift), we conclude that 
it is indeed a strong coupling region.
One easily sees that $(\zeta-\zeta_1) \sim (r-r_1)^{2/3}$, for some 
constant $r_1$ which we henceforth absorb into $r$ so that
the metric behaves like  
\be
ds^2 \simeq 
dr^2 + ({\rm const.})\ r^{2/3} \eta_{\m\n}dx^\m dx^\n \ ,\qq {\rm as}
\quad r\to 0^+\ .
\label{h1j}
\ee

Region 2 follows by expanding the fields around $b-a\zeta  \simeq 0$. 
If we let $\zeta_2= b/a$, we find the following behaviour
\be
e^{-\sqrt{2}\phi} \simeq {3 \over 2\sqrt{2} a^{2/3}}(\zeta_2-  
\zeta)^{1/3} ~, ~~~~ e^{\pm \omega} \simeq  {2\ov a^{1/3}}
{1 \over (\zeta_2 -\zeta)^{1/3}} ~, ~~~~ {\rm as} ~~
\zeta \rightarrow \zeta_2^- ~. 
\label{paraaa}
\ee
In this vicinity we have $\omega \rightarrow 
\pm \infty$, depending on the branch in the upper or lower $\omega$  
half-plane, and since $\phi \rightarrow \infty$ we also 
have $\chi \rightarrow 
+\infty$, which indeed describes the strong coupling region 2 as 
advertized. 
Using $(\zeta_2 -\zeta)\sim (r_2-r)^{6/7}$, for some positive $r_2$,
one finds that 
\be
ds^2 \simeq 
dr^2 + ({\rm const.})\ (r_2-r)^{2/7} \eta_{\m\n}dx^\m dx^\n \ ,\qq {\rm
as}
\quad r\to r_2^-\ .
\label{h2j}
\ee

Region 3 arises by first letting $\phi \rightarrow - \infty$. Then,
the heterotic equations simplify in this limit to the following
system:
\ba
{d\phi \over d\zeta} &=& -\sqrt{2} {d\chi \over d\zeta} = {1 \over 4}
e^{-\sqrt{2}\phi} {\rm sinh}^2 \omega ~, 
\nonumber\\
{d\omega \over d\zeta} &=&- {1 \over 2\sqrt{2}} e^{-\sqrt{2}\phi} 
{\rm sinh}\omega {\rm cosh}\omega ~.  
\ea
Clearly, $\sqrt{2} \chi = - \phi $ (up to a constant) and so 
this is also a strong coupling region. Since we have 
$\phi \rightarrow - \infty$,
the dominant dependence on $\zeta$ is easy to extract. We find
\be
e^{\sqrt{2}\phi} \simeq \left({3 \over 2c^2\sqrt{2}}\right)^{1/3} 
(\zeta - \zeta_3)^{1/3} ~, ~~~~ e^{\pm \omega} \simeq 2 
\left({2\sqrt{2} \over 3c}\right)^{1/3} {1 \over (\zeta - 
\zeta_3)^{1/3}} ~, ~~~ {\rm as} ~~ \zeta \rightarrow \zeta_3^+ ~,  
\label{paraac} 
\ee
where $c$ and $\zeta_3$ are constants of integration. In this case
we see that $\omega \rightarrow \pm \infty$ as it was initially 
stated. Finally, $\zeta-\zeta_3 \sim (r-r_3)^{6/7}$ for some constant 
$r_3$ which we again absorb into $r$. The behaviour of the metric near
$r=0$ is 
\be
ds^2 \simeq 
dr^2 + ({\rm const.})\ r^{2/7} \eta_{\m\n}dx^\m dx^\n \ ,\qq {\rm as}
\quad r\to 0^+\ .
\label{h2j3}
\ee

\begin{figure}[ht]
\begin{center}
{\scalebox{.55}{\includegraphics{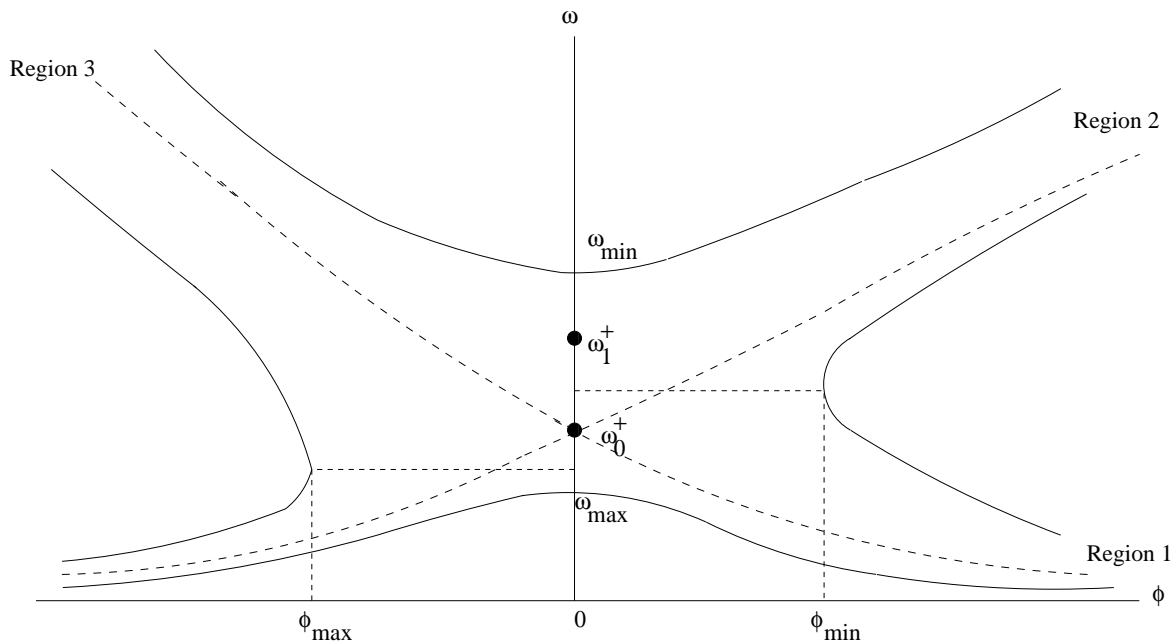}}}
\caption{Qualitative behaviour of different trajectories in the
upper-half
$\phi$--$\om$ plane for the heterotic case.}
\end{center} 
\label{het}
\end{figure}

The behaviour of different trajectories in the $(\phi,\om)$ plane is 
depicted in Figure 3. Because of the symmetry $\om \to -\om$ we have 
restricted to windings $\om \geq 0$ without loss of generality,
in particular solutions that start in the upper-half plane cannot end 
somewhere in the lower half-plane. 
The qualitative characteristics of the various trajectories can be
deduced by
treating $\om$ (or $\phi$) as an independent variable and consider the 
differential equation for $d\phi/d\om$. 
Before we examine the 
different trajectories in some detail, note by inspection of 
the differential equation
$d\chi/d\om$ and from the strong coupling behaviour that we already
know,
that there can be trajectories along which the
string coupling $e^\chi$ develops a minimum
at a point in the $\phi$--$\om$ plane where
\be 
e^{2\sqrt{2} \phi}= {\sinh^2\om\ov \sinh^2\om-2}\qq \Rightarrow 
\qq \phi>0\quad {\rm and} \quad \om > \om_1^+ \equiv \ln(\sqrt{3}+\sqrt{2})\ .
\label{kh2}
\ee
For $\phi<0$ there is no such minimum and
the string coupling $e^\chi$ keeps increasing with increasing
$\om$.
Let us now discuss the different types of trajectories as 
depicted in Figure 3:

\no\underline{Trajectories with an extreme $\om$}: These are 
trajectories where for some $\om$ we have $\phi=0$ and hence
$d\om/d\phi=0$
showing that these trajectories intersect perpendicularly the axis $\phi
= 0$.
Since 
\be
{d^2\om\ov d\phi^2}\bigg|_{\phi=0}=\ {\sinh 2 \om\ov \sinh^2\om -1}\ ,
\label{fii1}
\ee
we have for $\om > \om_0^+=\ln(\sqrt{2}+1)$ ($ < \om_0^+)$ that this
$\om$ 
is a minimum (maximum) in 
the trajectory. Two such typical trajectories are drawn in Figure 3. The 
lower one connects the weak coupling region $(\phi,\om)=(-\infty,0)$ 
with the strong coupling region 1, after reaching a maximum for the
winding 
$\om_{\rm max}\in (0,\om_0^+)$. Since \eqn{kh2} can never be satisfied,
it can be easily seen that $d\chi/d\phi>0$
throughout the trajectory and hence the string coupling $e^\chi$ is 
progressively increasing with $\phi$.
Another such trajectory 
connects the strong coupling regions 2 and 3 
going through a minimum for the winding $\om_{\rm min}\in
(\om_0^+,\infty)$.
For this trajectory there is of course a minimum for the string coupling 
at a point described by \eqn{kh2}.
We also note that for the trajectories with $\om\leq \om_{\rm max}$
that connect the  weak coupling region to region 1,
the variable $\zeta$ and hence also $r$ 
decrease as $\phi$ increases, with $r$ going from $+\infty$ to 0. For
the 
trajectories with $\om\geq \om_{\rm min}$ that connect regions 3 and 2, 
$\zeta$ and $r$ increase as $\phi$ increases, with $r$ going from 0 
to some $r_2>0$. Note that the value of $r_2$ depends on the chosen
trajectory.

\no\underline{Trajectories with an extreme $\phi$}: There are
trajectories 
for which $d\phi/d\om=0$ and hence there is an extreme value for $\phi$.
We find that the condition for this is
\be 
e^{2\sqrt{2} \phi}= {\sinh^2\om\ov 2-\sinh^2\om}\qq \Rightarrow 
\qq 
\left\{\begin{array}{ll}
\phi>0\ , & \ {\rm for}\quad \om_0^+ < \om < \om_1^+ \ ,
\\
\phi <0 \ , & \ {\rm for}\quad 0< \om< \om_0^+   \ .
\end{array} 
\right.
\label{kh3}
\ee
Since 
\be
{d^2\phi\ov d\om^2}\bigg|_{{d\phi\ov d\om} =0}=\ {\sqrt{2}\ov \sinh^2\om
-1}\ ,
\label{fi21}
\ee
we have for $\phi >0$ that this $\phi$ is a minimum (indicated by
$\phi_{\rm min}$) occuring for $\om\in
(\om_0^+,\om_1^+)$, whereas for $\phi<0$ it is a maximum (indicated by
$\phi_{\rm max}$) occuring for $\om\in (0,\om_0^+)$. 
Two such typical trajectories are drawn in Figure 3. The one with
$\phi<0$
connects the weakly coupled region having $(\phi,\om)=(-\infty,0)$ 
with the strongly coupled
region 3. Since $\phi<0$ there is no mimimum for the string coupling,
which 
increases with $\om$ in accordance with \eqn{kh2}.
The trajectory with $\phi>0$ connects the two strongly coupled regions 1
and
2. For such trajectories, besides the depicted minimum value for $\phi$,
there is also a minimum value for the string coupling according to
\eqn{kh2}.
Note that for the trajectory with $\phi>0$ ($\phi<0$) 
the variables $\zeta$ and $r$ increase (decrease) as 
$\om$ increases, with $r$ ranging from 0 to some $r_2>0$ 
(from $\infty$ to 0). Again, the value of $r_2$ depends on the chosen
trajectory.

We also note that region 2 contains trajectories that come either 
from region 1 or from region 3.\footnote{In these regions we have the
linear 
behaviour $\sqrt{2}\phi \simeq \om +\ln\left(\sqrt{2} a/3\right)$ 
(for region 2) and 
$\sqrt{2}\phi \simeq -\om +\ln\left(2/c\right)$ (for region 3).  
For region 1 we have instead 
$\sqrt{2}\phi \simeq - 2 \ln\om - \ln (3/2\sqrt{2} a)$.}
The family of such trajectories is parametrized 
by the constant $a$ in \eqn{paraaa}. 
The values of $r_2$ and $a$ are clearly related, 
but we do not know exactly how.
There exists a critical value, which is 
numerically found to be $a_{\rm crit} \simeq 0.9 $, 
such that for $a > a_{\rm crit}$ ($a < a_{\rm crit}$) we 
connect to region 1 (3).
Similarly, in region 3 the different trajectories are parametrized by
the constant $c$ in \eqn{paraac} and there is also a 
critical value $c_{\rm crit}
\simeq 1.2$; 
for $c>c_{\rm crit}$ ($c<c_{\rm crit}$) these trajectories end up in the
weakly coupled region (region 2). Of course, for the trajectories that
connects
the regions 2 and 3 the constants $a$ and $c$ are related, but how
precisely, 
cannot be answered without knowledge of the explicit solution.

\no\underline{Trajectories ending at critical points}: These are the
trajectories depicted with the dashed lines and occur when the constants
$a$ 
and $c$ assume the critical values we mentioned. We emphasize that
despite 
appearances, these trajectories
have the critical points as their end-points, and they do not simply
pass
through them. In other words, they do not connect the strongly coupled
regions
1 and 3 nor the weakly coupled region $(\phi,\om)=(-\infty,0)$ 
with the strongly coupled region 2. The critical point indicated in
Figure 3,
represents a weak coupling point for trajectories that come either 
from region 1 or from region 3. In these cases the solution in the
vicinity 
of the critical point is given by \eqn{ght1} and the variables $\zeta$ 
or $r$ increase to $+\infty$ as we approach the critical point.
In contrast, for trajectories that go either to
region 2 or to the weakly coupled region with $(\phi,\om)=(-\infty,0)$,
this critical point is at strong coupling. Then, the solution in its
vicinity 
is given by \eqn{ght2} and the variables $\zeta$ or $r$ increase
as we get away from the critical point. 
We note that the trajectory going to region 2 connects 
strong coupling to strong coupling so that $\chi$ must have a minimum
along 
its way. This is consistent with the fact that  
such a minimum occurs if and only if $\phi>0$ and 
$\om>\om_1^+$. This trajectory is the only one among the 
four for which $r$ takes values in a finite interval $[0,r_2]$.

In section 6, we will discuss certain criteria based
on the propagation of a  free quantum particle on the corresponding 
backgrounds that  render some of these trajectories unphysical.

We conclude the discussion of the heterotic equations 
by mentioning the change of variables 
\be
x={\rm tanh} \omega ~, ~~~~~ y={\rm cosh}(\sqrt{2}\phi) 
{{\rm sinh}\omega \over {\rm cosh}^2 \omega} ~, 
\label{newpp}
\ee
which cast the differential equations
for $\phi(\omega)$ into a simpler looking 
form for $y(x)$ (after taking its square)   
\be
x^2(y^{\prime})^2 - y^2 = x^2(x^2 - 1) ~. 
\label{gh3}
\ee
Unfortunately, we are still lacking its general solution 
in analytic form. Nevertheless, the
form of the eq. \eqn{gh3} enables 
to perform a systematic study of the perturbative 
expansion around the critical points \eqn{crii}, which 
are located at $(x,y)= \pm \left({1\ov \sqrt{2}},\ha\right)$
in the parametrization \eqn{newpp}. 
We have the following perturbative expansion, with 
the upper (lower) signs  
refering to the corresponding signs in \eqn{crii}, 
\be
y=\pm \ha + \sum_{n=0}^\infty a_n 
\left(x\mp {1\ov \sqrt{2}}\right)^{n+2}\ ,
\label{pp3}
\ee
where the coefficient $a_0$ of the expansion is given by 
\be
2 a_0^2 \mp a_0 -2 =0 \ . 
\label{ad1}
\ee
We also have the following recursive relations for $n\geq 1$: 
\ba 
\left(2 (n+2) a_0 \mp 1\right) a_n & =& \pm 2\sqrt{2} \d_{n,1} -
\d_{n,2} 
-\ha \sum_{k=1}^{n-1} (k+2)(n-k+2) a_k a_{n-k} 
\nonumber\\
&& \mp \sqrt{2} \sum_{k=0}^{n-1} (k+2)(n-k+1) a_k a_{n-k-1} 
\label{ad2}\\
&& - \sum_{k=0}^{n-2} ((k+2)(n-k)-1) a_k a_{n-k-2} \ .
\nonumber
\ea
Note that the freedom to choose the constant $a_0$ in \eqn{ad1}
corresponds to the two independent solutions of \eqn{gh3} that arise
from taking the square root. Since the constant $a_0$ determines the 
sign of $y^\prime$, we have 
for $\om\ge 0$ that $a_0>0$ for $\phi>0$, whereas $a_0<0$ for 
$\phi<0$.
The first few terms of the expansion are given by 
\be 
y = \pm \ha + a_0 \left(x\mp {1\ov \sqrt{2}}\right)^2 
- {2 \sqrt{2} (a_0\pm1)\ov 6a_0 \mp1}
\left(x\mp {1\ov \sqrt{2}}\right)^3 + \cdots\ .
\ee
One could develop a similar perturbative expansion
around any of the strongly coupled regions 1, 2 and 3 or
in the vicinity of the weakly coupled region
$(-\infty,0)$. The corresponding points, in the $x$--$y$ plane, around 
which one should expand are $(x,y)=
(0,+\infty)$, $(1,\sqrt{2} a/3)$, $(1,c/2)$ and $(0,0)$ respectively.

Finally, let us comment on the possibility to have a compact 
coordinate $r$ as suggested by the microcanonical analysis of 
refs. \cite{bv, neil, rob}. This would imply that all fields must 
come back to their initial value after some period in $r$. 
We have seen that there are two types of trajectories. On the one 
hand, we have those for which $r$ takes values in a semi-infinite 
interval. These are the trajectories that connect the weak-coupling 
region $(\phi,\om)=(-\infty,0)$ to any other region, as well as 
those connecting regions 3 or 1 to the critical point. Clearly, 
these solutions are not, and cannot be made periodic in $r$. 
On the other hand, we have the trajectories ending in region 2 
for which $r$ takes values in the finite intervals $[0,r_2]$
with the value of $r_2$ depending on the chosen trajectory. 
This is crucially different from what we found in the type-II cases. 
Note that trajectories in this second class always go from strong 
to strong coupling. It may well be that these solutions can be 
continued into periodic functions of $r$ beyond 
these intervals, tracing the corresponding trajectories back and forth. 
The situation is examplified by the parabola $y=x^2$ in the $x - y$
plane
with $x$ a function of $r$ given by
$x(r)=\ln {1-\cos r\over 1+\cos r}$. This clearly is a periodic 
function in $r$ and hence $r$ can be taken as a compact coordinate with 
values on the circle $[0,2\pi]$. On the other hand, if we only 
look at the interval $r\in [0,\pi]$, $x(r)$ goes  from 
$-\infty$ to $+\infty$  and 
we trace the parabola just once. This is somewhat similar 
to our trajectories going into region 2. 
Without an 
explicit solution at our disposal, it is probably impossible to 
decide whether or not we can continue these solutions into periodic 
ones or not.

\section{Physical boundary conditions}
\setcounter{equation}{0}

Next, we examine the boundary conditions which are physically relevant
for the domain wall backgrounds. Note that until now the integration 
constants of the corresponding first order equations were left 
undetermined, and so any solution could be useful mathematically,
as there is no preference among the different orbits in the parameter
space, say $(\phi, \omega)$. We now apply
some general criteria for restricting the 
physical range of the parameters
within the context of supergravity using the propagation of a quantum
test particle on the corresponding backgrounds. Then, we discuss the 
criteria that string theory may impose, although the situation is 
less clear in that case.

Curvature singularities lead to singular classical dynamics for 
test particles. If they persist at the quantum level,
the theory is considered as unphysical.
Unfortunately, we do not know enough about string theory in such
backgrounds
in order to answer this question definitely.
Instead, we can ask the simpler question whether the 
propagation of a quantum test particle is
well defined in the presence of a curvature singularity.
A singular classical propagation, as indicated by an incomplete
geodesic motion, does not necessarily lead to a singular wave
propagation 
since the space-time can produce an effective barrier that 
shields the classical singularity. In general terms, a singularity will
be 
physical if the evolution of any state is uniquely defined for all
times.
The background is then called wave-regular. This criterion 
is identical to finding a unique
self-adjoint  extension of the wave-operator \cite{wald, homa,
ishibashi}.
This can be quickly investigated by looking at the two solutions of the
wave equation locally near the singularity. 
Note that the wave equation arises, for instance,
when one considers linearized
graviton fluctuations along the three-dimensional brane embedded into
the 
four-dimensional domain wall ansatz \eqn{metriki}, i.e., $\eta_{\m\n}\to 
\eta_{\m\n} + h_{\m\n}$.\footnote{ This has been shown in 
the context of the AdS/CFT correspondence and in warped factor 
compactifications in \cite{BrS2} (for an exhaustive exposition see 
\cite{DeWotsei, aft}). The proof in the present case can also be given 
along similar lines.}
In such cases, in addition to having a unique 
evolution of initial data, one has to ensure that these fluctuations
never 
become strong, which would invalidate the linearized approximation. 

A remark is in order. Since we are at finite temperature we have compact 
euclidean time and it is not clear right away what is meant by dynamics 
and evolution. Our  coordinates  $r$ and $x^\mu$ with $\mu=1,2,3$ are
all spatial.
To study the ``adiabatic" dynamics of a single test particle one 
should add a test-particle time coordinate with $g_{tt}=1,\
g_{t\mu}=g_{tr}=0$. 
Indeed, a {\it single} test particle only propagates in the thermal
background
specified by the  metric but does not interact with the thermodynamical
system.
Thus in the following, we use the five coordinates:  test-particle
time $t$ and four 
space coordinates $x^i$ split into $r$ and $x^\mu,\ \mu=1,2,3$.

The relevant part of the wave equation is 
\be
\frac{d}{dr} \left( \sqrt{-g} g^{rr} \frac{d\psi}{dr}\right) =0\ .
\label{wwavv}
\ee
We will soon justify the need to keep only this term in the wave
equation
for our class of examples, where the
metric behaves in general as in \eqn{hjhj3} below.
In order to check the normalizability of the two independent solutions
of the wave equation, we will use the Sobolev norm 
\be 
\frac{q^2}{2} \int_\S \sqrt{-g} g^{tt} \psi^\dagger \psi +
\frac{1}{2}\int_\S \sqrt{-g} g^{ij} \del_i\psi^\dagger \del_j \psi\ ,
\label{gfh2}
\ee
where $q$ is a constant and the integrals are 
performed on a constant-time
hypersurface $\S$.
The Sobolev norm is bounded from above by 
the energy of the fluctuation associated with the wave, and
guarantees that the back-reaction of the fluctuation
is indeed small \cite{ishibashi}. According to this
criterion there exists a unique self-adjoint extension of the
wave-operator 
and a unique evolution of the system, 
thus rendering the space wave-regular, 
if {\it only one} of the two independent solutions of \eqn{wwavv} is
normalizable 
near the singularity. This solution is then kept, whereas the other 
is discarded. We should note that this criterion is non-trivial for
time-like singularities. For null singularities the space-time is
globally 
hyperbolic and there should be a
unique self-adjoint extension of the wave-operator since the evolution 
is ordinary Cauchy, and hence unique in this case.

The choice of Sobolev norm is compatible in our cases
with an alternative criterion,
namely that in geodesically incomplete space-times there should be no
leak of conserved quantities through the singularity.
For the translational isometries 
that remain unbroken by our metric ansatz \eqn{metriki}, 
this amounts to the condition (see, for instance, \cite{Cohen})
\be
\sqrt{-g} g^{rr} \del_i \psi \del_r \psi = 0  
\label{hfg11}
\ee
at the singularity. In order to have 
a unique evolution of the system, only one of the two independent 
solutions of eq. \eqn{wwavv}
should satisfy \eqn{hfg11}. For the other solution, which is discarded, 
the right hand side of \eqn{hfg11} should diverge.

Returning to our metrics, we have seen that they develop a curvature
singularity as $r\to 0^+$ or $r\to r_2^-$. We will discuss the case 
of a singularity at $r=0^+$, the other case being analogous.
Then the asymptotic behaviour of the metric is of the form
\be
ds^2 \simeq 
dr^2 + ({\rm const.})\ r^{2\n} \eta_{\m\rho}dx^\m dx^\rho \ ,\qq {\rm
as}
\quad r\to 0^+\ ,
\label{hjhj3}
\ee
with $\n$ being a positive constant that can be read  
from the appropriate expressions in sections 4 and 5.
In the case of the strongly coupled solution \eqn{ght2} around the
critical point $(\phi,\om)=(0,\om_0^+)$, we have that $\n=1$, and 
for all other cases $0<\nu<1$.
Applying the general formalism for a background 
behaving as in \eqn{hjhj3} we first see that the two independent
solutions 
of \eqn{wwavv} are $\psi_1\sim 1$ and $\psi_2 \sim r^{1-3\n}$ (or 
$\psi_2\sim \ln r$ if $\n=1/3$). Note that if we had 
treated the full wave equation
in \eqn{wwavv}, we would have added a term proportional to $r^\nu\psi$
on the
left-hand side. Such term perturbs the two independent solutions 
that we have just 
mentioned by a term proportional to $r^{2-2\nu}$ and $r^{3-5\n}$, 
respectively. Therefore, near the singularity 
occuring at $r=0$, this extra term 
can be neglected if $\n<1$, or does not change the power of $r$ if
$\nu=1$.

Applying either one of the above criteria, we find that backgrounds 
with a single singularity 
behaving as in \eqn{hjhj3} are wave-regular
for $\n\ge 1/3$.\footnote{This condition also guarantees that the
solution 
$\psi_2\sim r^{1-3 \n}$ that diverges at the singulariry occuring at
$r=0$, 
is discarded.
Hence, if the wave equation that we are examining corresponds to
linearized 
graviton fluctuations, these fluctuations will indeed remain small,
thus not invalidating the approximation.}
Hence, for the genuine 
type-II, as well as for the hybrid type-II solutions, the backgrounds
are
wave-regular when the constants $c$ and $b$, respectively,
are negative or zero, while positive $c$ or $b$ are unphysical. 

For the heterotic case, we have to distinguish solutions with 
semi-infinite intervals of $r$ (those not going into region 2) 
and solutions with finite intervals of $r$ (those going into region 2). 
For the former there is only one singularity at $r=0^+$ and we find that
all solutions that contain in some limit the region 3 do not satisfy
the criteria, and hence are unphysical.
The only physical solutions in this class are those corresponding 
to the trajectories 
that connect region 1 to the weak coupling region, or the
special trajectory that go from the 
critical point $\om_0^+$ to region 1 or to the weak coupling region. 
Note that for all these trajectories $\om$ 
remains bounded by $\om_0^+$.
Now consider the solutions that have a finite interval of $r$. 
They have two singularities, one at $r=0^+$ and one at $r_2^-$. 
If at each singularity one of the two solutions to the wave equation 
is discarded, no solution will be left, as in general the solution
discarded 
at one singularity will not be the same as the one discarded at the
other.
Such a 
background can be wave-regular only if one of the singularities 
discards one solution and the other singularity does not impose 
any condition, i.e. we need $\nu\ge 1/3$ at one singularity and 
$\nu < 1/3$ at the other. We have $\nu = 1/3,\ 1/7,\ 1/7$ and $1$ 
for regions 1, 2, 3 and the critical point.
We conclude that solutions corresponding 
to trajectories from region 2 to region 1 or  
to the critical point are wave regular. The solution 
corresponding to the trajectory from region 2 to region 3 
however is not wave-regular.
It is evident that a better understanding of the string rather than the 
particle propagation is desirable on such  
backgrounds.\footnote{The above criteria have been succesfully applied 
within the AdS/CFT correspondence in many backgrounds related to the 
Coulomb branch of ${\cal N}=4$ and ${\cal N}=2$ super Yang--Mills
theories
\cite{BrSf}.
In those cases the lack of information on string theory beyond the 
supergravity approximation is compensated by the information related to 
gauge theory expectations.}

From the string thermodynamic point of view it has been 
argued \cite{bv, neil, rob} that  the exponential growth of the number 
of states forces one to work in the microcanonical ensemble, in which
case  
the specific heat may turn negative unless {\it all} spatial 
dimensions are compact; of course some compact dimensions may 
well be large (see also \cite{tan, djnt}). 
Within our domain wall ansatz, compactness or not 
of the space coordinates $x^\mu$  is not an issue since the 
solutions do not depend on those directions. However, compactness of the
$r$ 
coordinate is possible only if the solutions $\phi(r), \chi(r)$ 
and $\om(r)$ at some $r$ equal their values at some other $r$. 
As we discussed in the previous section, this is not the case 
for the type-II solutions. For the heterotic case we have some 
solutions that are parametrized by a finite interval in $r$. 
However, without the explicit solution at our disposal, it is 
not clear whether they can be continued to periodic solutions or not. 
In any case, before 
jumping into conclusions about  our solutions being physical or
unphysical
within string theory, one should bear in mind that  the 
compactness criterion was established within the microcanonical 
treatment of string theory, which assumes that the theory remains
weakly coupled everywhere.\footnote{
It is somewhat ironical that the only solutions that perhaps 
could be continued periodically actually are never weakly coupled.
} 
It is clear from our solutions that they 
always contain regions of arbitrarily strong coupling, thus 
invalidating the assumptions that could sentence some or all of them  
unphysical. 

Another argument in favour of our solutions is 
their supersymmetry. As shown above, all solutions 
of the first order equations preserve half of the supersymmetries,
by their construction,  
while the standard high-temperature phase of strings breaks all 
supersymmetries. Clearly, at this point we cannot say anything 
further as definite
about the relevance and physical properties of our different domain wall
solutions within string thermodynamics. Ideally, a string inspired
criterion
would render physical all or a proper subset of those
backgrounds, which have already been named physical 
using the criteria of supergravity in this section.

\section{Conclusions and discussion}

In this paper we have constructed domain wall solutions of the
effective supergravity which describes all possible 
high-temperature phases of the known $N=4$ superstrings. We have
used a universal thermal potential that contains the $s$, $t$, $u$ 
moduli and their couplings to the three (lightest) winding modes, 
which can become tachyonic above the Hagedorn temperature, thus
triggering the thermal instabilities. Our solutions contain 
non-trivial winding fields, which vary with respect to the domain
wall variable $r$, and exhibit a common property in that they 
always extend to regions of strong coupling. We have presented the exact 
solutions for the type-IIA and type-IIB sectors of the theory
and a certain
(self-dual) hybrid sector, and investigated the general structure of
the solutions for the heterotic sector by extracting their behaviour
in the vicinity of the weak and the strong coupling points. From a 
geometrical point of view we found that none of the type-II solutions
has
support on a compact spatial dimension, 
namely the direction parametrized by the
variable $r$. For the heterotic case, the same is true for all solutions
that contain a weakly coupled region, while solutions that never are 
weakly coupled have support only on finite intervals in the variable
$r$.
Whether or not these latter solutions can be continued into periodic
functions
of $r$ could not be answered without having an explicit solution 
at our disposal.
From a supergravity point of view, for each type of theory, there are
subclasses of domain wall backgrounds which are selected by imposing 
specific boundary conditions that lead to consistent propagation of
a quantum test particle. 

The main property of the domain wall solutions is their supersymmetry; 
they all satisfy first order differential equations, which arise as BPS
conditions for gravitational backgrounds that preserve 1/2 of the 
supersymmetries. The supersymmetry of our solutions is the key to 
understand why, for all of them, tachyonic instabilities never occur,  
even though the temperature may become arbitrarily high.
Of course, we are only considering solutions of an effective
supergravity
and not of the full string theory, but it may well be that they 
point to a new finite-temperature phase of superstrings 
which is supersymmetric and has no thermal instabilities, i.e., no
Hagedorn 
temperature. One may then speculate that, as superstrings are heated 
up from zero temperature, they prefer to go into this more symmetric
phase 
which is stable due to supersymmetry, and that the ``ordinary" 
high-temperature phase with the Hagedorn instability is never reached.

Let us now address a few points one might object to our solutions.
First, one might worry about the status of the domain
wall ansatz in theories of gravity coupled to scalar fields and having a
potential with run-away directions that account for the usual
instabilities 
at the Hagedorn temperature. Note however, that this is not a priori 
forbidden by any general considerations alone. Recall 
that domain wall
solutions with infinite tension (corresponding to infinite
central charge of the supersymmetry algebra) 
have already found numerous applications
in supersymmetric gauge theories, most notably in supersymmetric quantum 
chromodynamics with one massless flavor,  
where one has models with run-away vacua. In fact, these are models with
rigid supersymmetry which admit stable field configurations that restore
one half of supersymmetry, as opposed to the stable but
non-supersymmetric
ground state, and which are characterized by constant positive energy
density \cite{dvali}.The domain wall solutions that we have constructed 
in supergravity 
for the problem of string thermodynamics could be viewed as the
gravitational
analogue of such supersymmetric configurations, having their own 
characteristic properties. Their stability 
is guaranteed by the saturation of the BPS bound
and can be further supported by analysing the spectrum 
of the graviton fluctuations on the domain wall backgrounds. Based on
general principles, one can show that the graviton spectrum is obtained
by 
computing the energy levels of an
equivalent non-relativistic problem in supersymmetric quantum mechanics
with a Schr\"odinger potential that is determined 
entirely by the conformal factor of the
metric. The details are not important as the only relevant point 
here is that the
spectrum is bounded from below (by zero) by the general properties of 
supersymmetric quantum mechanics, thus rendering a physical spectrum
for the graviton fluctuations on the supersymmetric domain 
wall backgrounds that preserve three-dimensional Poincare invariance.  

Second, one should also study metric fluctuations taking us
away from the class of conformally flat metrics of the domain wall
backgrounds. We have not investigated this issue, but we expect that 
an analysis along the line that can e.g. be found in \cite{DeWotsei,aft} 
is possible.

Third, a well-known instability  of hot
flat space was  analysed in \cite{gross}. In this simplest 
example it was found that the classical Jeans instabilities arise as a
tachyon 
in the graviton propagator, using small fluctuations about hot flat
space,
and it was further suggested that hot flat space will nucleate black
holes.
Inevitably, this is the fate of gravitational systems due to the
attractive nature of gravitation. In the case of a relativistic medium, 
the Jeans instability implies that a thermal ensemble of sufficiently 
large volume will collapse into a black hole. Of course, it remains to
be
seen how these results generalize to domain wall backgrounds of our 
effective supergravity by performing calculations as in ordinary quantum
gravity. One might fear that most of our solutions will be afflicted by
Jeans
instabilities because they do not have support in small volumes.
However, their defining BPS property ensures their stability quantum 
mechanically as well. Put differently, the BPS equation which is
interpreted
as a ``no force" condition, although this interpretation 
is more appropriate for 
asymptotically flat backgrounds with a definite Newtonian limit, ensures
that no gravitational collapse will occur.    

Clearly, spherically symmetric black hole type solutions of the
effective 
supergravity have to be analysed in the future before addressing the 
long standing problems of quantum gravity within string thermodynamics. 
These general remarks put in perspective  
future attempts to construct other non-trivial
solutions of the effective supergravity for the $s$, $t$, $u$ moduli and
the
string winding modes beyond the supersymmetric domain wall ansatz.
An important lesson that we already learned, even from
the simplest solutions we have constructed here, is that they
cannot be entirely confined in regions of weak coupling only.

It would be interesting to study the
most general domain wall solution of the effective supergravity that 
describes all possible phases of $N=4$ superstrings at finite
temperature
and includes all six scalar fields.
The resulting system of first order non-linear differential equations 
does not seem  tractable by analytical methods alone,  but a combination 
of numerical analysis and analytical calculations around certain 
points may well provide a reasonable global picture, similar to what we 
did in the heterotic case.
In the appendix, below, 
we summarize for completeness some general remarks about
the search for periodic orbits
in order to appreciate the difficulty to establish analytic criteria for 
their existence in the general case. 

Finally, it will also be interesting to study 
the higher dimensional interpretation
of our domain wall configurations, as in other theories of gauged 
supergravity. This might be related to thermodynamics of string 
theory in D-brane backgrounds as has already been discussed in 
the weak-coupling limit in \cite{abkr, afian}. One may also try 
to extend the formalism to construct domain 
wall junctions.

\bigskip\bigskip
\centerline{\bf Acknowledgements}
\noindent
One of us (I.B.) wishes to thank the members of the Institute of Physics 
in the University of Neuch\^atel for the warm hospitality extended to
him
during a visit and the collaboration that initiated the present work. 
We also thank G. Katsiaris for his assistance with the numerical aspects
of the heterotic sector. This research was supported
by the Swiss National Science Foundation, the European Union under
contracts 
TMR-ERBFMRX-CT96-0045 and -0090, by the Swiss Office for Education and
Science and by RTN contract HPRN-CT-2000-00122.
   
\newpage

\appendix
\section{Remarks on periodic orbits}
\setcounter{equation}{0}

Although we have seen quite explicitly that all our type-II solutions 
are non-periodic in the domain wall coordinate $r$, and the same
probably 
is true also for the heterotic solutions, it is much less clear
whether this will still be true for the solutions of the full set of six 
equations \eqn{fi1} - \eqn{hh3}. Since for the latter one probably has 
to rely heavily on numerical methods, it would be nice to have at least 
some analytical tools at one's disposal. One possibility is to try to 
establish the existence or non-existence of periodic orbits in the 
solution space of our non-linear system of 
first order differential equations for the domain wall configurations.
Clearly if there are such periodic orbits in the solution space 
(e.g. in the $\phi-\om$ plane in the heterotic case) then
the corresponding solution will also be periodic in $r$. Note however
that
the converse is not necessarily true as would be examplified by the 
heterotic trajectories going into region 2, should it turn out that they 
can be periodically continued in $r$. Of course, then one might still
consider 
them as periodic orbits in the $\phi-\om$ plane, although degenerate
ones.
Finally, for reasons already discussed at 
length in the introduction and discussion section, it is not clear
whether 
periodicity may serve as a criterion within
string thermodynamics to distinguish physical from unphysical solutions.

It is well known from the theory of dynamical systems that the existence
of periodic solutions, i.e., closed loops in the two-dimensional
parameter 
space $(\phi, \omega)$, of a first order system 
$\dot{\phi} = P(\phi, \omega)$ and 
$\dot{\omega} = Q(\phi, \omega)$ is a very important and delicate
question
that often can be established only numerically (unless the general
solution
is known in closed form). The boundary conditions select specific orbits
$\phi(\omega)$ by fixing the integration constants, and so by singling
out
the periodic orbits, if there are any among the solutions, corresponds
to
specifying some physical boundary conditions. 

To appreciate the difficulty in establishing the existence of periodic
orbits analytically, we recall briefly some basic elements of 
Poincar\'e's theory 
(see, for instance, \cite{book} for an elementary account). There, one
has the notion of a limit cycle, i.e. a stable closed curve in the
parameter
space, independent of initial conditions, towards which solutions tend
in
an asymptotic sense, or from which they unwind, as it were, as 
$\zeta \rightarrow \pm \infty$. A well known theorem states that if a
limit
cycle exists in a given system, then the existence of periodic orbits is
guaranteed, even though their explicit construction could be a 
difficult task. On the other hand we have Bendixson's theorem stating
that
if one considers the function  
\be
J = {\partial P \over \partial \phi} + {\partial Q \over \partial
\omega} 
\ee
in any given domain in $(\phi, \omega)$ which is bounded by a simple
curve 
${\cal C}$, the system will have no limit cycles inside that region if
$J$ has constant sign in it. Therefore, all we can learn on general 
grounds is for which regions there are no limit cycles; if $J$ changes 
sign in a bounded region (and hence it is bound to vanish somewhere in
it), 
the system will not necessarily have limit cycles in it, but
there is a good chance that it will. Furthermore, there
are systems with periodic orbits that have no limit cycles. 

According to these results, we can already
see a difference between the type-II and heterotic sectors concerning 
the possibility to have (or not to have) 
limit cycles in their parameter space. 
Computing
Bendixson's function for these two sectors (using the coordinate 
$\zeta$ that decouples the $\chi$-dependence from the flows of the
fields $\phi$ and $\omega$) one finds the result
\ba
J_{{\rm II}} &=& -{1 \over 2\sqrt{2}} \left(2e^{\sqrt{2} \phi} + 
e^{-\sqrt{2} \phi} (1 + 3{\rm sinh}^2 \omega)\right) , 
\nonumber\\
J_{{\rm hyb}} &=& -{1 \over 2\sqrt{2}} \left(2e^{\sqrt{2} \phi} + 
1 + 2{\rm sinh}^2 \omega \right) , 
\\
J_{{\rm het}} &=& -{1 \over \sqrt{2}} \left({\rm cosh}(\sqrt{2}\phi) 
-3{\rm sinh}(\sqrt{2}\phi){\rm sinh}^2 \omega \right) . 
\nonumber
\ea
It is obvious that $J_{{\rm II}}$ and $J_{{\rm hyb}}$ are strictly
negative
everywhere in the parameter space $(\phi, \omega)$, 
and so according to Bendixson's theorem we learn that there 
are no limit cycles in these cases. Of course, this by itself does not
rule out the existence of periodic orbits, but we already know 
from our explicit solutions that
for the pure type-II and hybrid type-II sectors there are none. 
On the other hand, the heterotic
sector could support limit cycles, since $J_{{\rm het}}$ vanishes along
the curve 
\be
{\rm coth}(\sqrt{2} \phi) = 3 {\rm sinh}^2 \omega ~. 
\ee
This curve has two branches, one in the upper $(\phi, \omega)$
half-plane 
and the other in the lower, which are related to each other by the
discrete symmetry $\omega \rightarrow - \omega$. Drawing the curve in
the upper half-plane we see that it starts asymptotically from $\omega = 
+\infty$ at $\phi = 0$ and drops monotonically to the value 
$\omega = \ln \sqrt{3}< \om_0^+$, which is approached asymptotically as 
$\phi \rightarrow + \infty$. 
Therefore, if one considers bounded domains in the parameter space
which intersect with 
Bendixson's curve $J_{{\rm het}} = 0$, they will potentially
contain limit cycles, although their existence is not at all guaranteed. 
Note that this curve is crossed precisely by the trajectories going 
to region 2. Of course, these trajectories are not contained in any 
bounded domain. Nevertheless, this is
still suggestive that maybe these trajectories can indeed be
periodically 
continued.   

It will be interesting to look for
closed trajectories in the higher 
dimensional parameter space with all the relevant fields turned on.



\newpage
\begin{thebibliography}{99}




\bibitem{strom} A. Strominger, ``Massless black holes and conifolds in 
string theory", Nucl. Phys. {\bf B451} (1995) 96, {\tt hep-th/9504090}.

\bibitem{vo} H. Ooguri and C. Vafa, ``Two-dimensional black holes and
singularities of CY manifolds", Nucl. Phys. {\bf 463} (1996) 55, 
{\tt hep-th/9511164}, and
``Summing up D instantons", Phys. Rev. Lett. {\bf 77} (1996) 3296,
{\tt hep-th/9608079}.

\bibitem{hag} R. Hagedorn, ``Statistical thermodynamics of strong
interactions
at high-energies", Nuovo Cimento Suppl. {\bf 3} (1965) 147.

\bibitem{ahag1} S. Fubini and G. Veneziano, 
``Level structure of dual resonance models",
Nuovo Cimento {\bf 64A} (1969) 811.

\bibitem{huang} 
K. Huang and S. Weinberg, ``Ultimate temperature and the early
universe",
Phys. Rev. Lett. {\bf 25} (1970) 895. 

\bibitem{ahag2} S. Frautschi, ``Statistical bootstrap model of hadrons", 
Phys. Rev. {\bf D3} (1971) 2821.

\bibitem{carlitz} R. Carlitz, ``Hadronic matter at
high density", Phys. Rev. {\bf D5} (1972) 3231.

\bibitem{mark} M. Bowick and L. Wijewardhana, ``Superstrings at finite
temperature", Phys. Rev. Lett. {\bf 54} (1985) 2485.

\bibitem{sundborg}
B. Sundborg, ``Thermodynamics of superstrings at high-energy densities", 
Nucl. Phys. {\bf B254} (1985) 583.

\bibitem{salomonson} P. Salomonson and 
B. Skagerstam, ``On superdense superstring gases: a heretic string model
approach", Nucl. Phys. {\bf B268} (1986) 349.

\bibitem{alvarez} E. Alvarez, 
``Strings at finite temperature", Nucl. Phys. {\bf B269} (1986) 596.   

\bibitem{sath} B. Sathiapalan, ``Vortices on the string world 
sheet and constraints on toral compactification", Phys. Rev. 
{\bf D35} (1987) 3277.

\bibitem{ian} Ya. Kogan, ``Vortices on the 
world sheet of a string; critical dynamics", JETP Lett. 
{\bf 45} (1987) 709.

\bibitem{aw} J. Atick and E. Witten, ``The Hagedorn transition 
and the number of degrees of freedom of string theory", 
Nucl. Phys. {\bf B310} (1988) 291.

\bibitem{bv} R. Brandenberger and C. Vafa, ``Superstrings in the 
early universe", Nucl. Phys. {\bf B316} (1989) 391.

\bibitem{neil} D. Mitchell and N. Turok, ``Statistical properties of
cosmic strings", Nucl. Phys. {\bf B294} (1987) 1138.

\bibitem{rob} S. Alexander, R. Brandenberger and D. Easson, ``Brane
gases
in the early universe", {\tt hep-th/0005212}.

\bibitem{ak} I. Antoniadis and C. Kounnas, ``Superstring phase 
transitions at high temperature", Phys. Lett. {\bf B261} 
(1991) 369.

\bibitem{adk} I. Antoniadis, J.-P. Derendinger and C. Kounnas, 
``Non-perturbative temperature instabilities in $N=4$ strings", 
Nucl. Phys. {\bf B551} (1999) 41, {\tt hep-th/9902032}. 


\bibitem{3d} E. Witten, ``Is supersymmetry really broken?", Int. J. Mod. 
Phys. {\bf A10} (1995) 1247,
{\tt hep-th/9409111} and
``Strong coupling and the cosmological 
constant", Mod. Phys. Lett. {\bf A10} (1995) 2153,
{\tt hep-th/9506101}.



\bibitem{adk2}
I. Antoniadis, J.P. Derendinger and C. Kounnas,
``Non-perturbative supersymmetry breaking and finite-temperature  
instabilities in N = 4 superstrings,'' {\tt hep-th/9908137}.
Proceedings of 6th Hellenic School and Workshop on Elementary Particle 
Physics: Corfu, Greece, 6-26 Sep. 1998. 


\bibitem{kr}
C. Kounnas and B. Rostand, 
``Coordinate Dependent Compactifications and Discrete Symmetries'',
Nucl. Phys, {\bf B341} (1990) 641.


\bibitem{N=4BPS}
J.H. Schwarz and A. Sen, ``Duality symmetries of 4-D heterotic strings",
Phys. Lett. {\bf B312} (1993) 105, {\tt hep-th/9305185} 
and ``Duality symmetric actions",
Nucl. Phys. {\bf B411} (1994) 35, {\tt hep-th/9304154}.

\bibitem{duff}
M.J. Duff, J.T. Liu and J. Rahmfeld, 
``Four-dimensional string-string-string triality",
Nucl. Phys. {\bf B459} (1996) 125,
{\tt hep-th/9508094}.

\bibitem{cvetic} 
M. Cvetic and D. Youm, ``Dyonic BPS saturated black holes of 
heterotic string on a six torus",
Phys. Rev. {\bf D53} (1996) R584, {\tt hep-th/9507090}
and ``Singular BPS saturated states and enhanced symmetries 
of four-dimensional $N=4$ supersymmetric vacua"
Phys. Lett. {\bf B359} (1995) 87, {\tt hep-th/9507160}.

\bibitem{cardoso}
G.L. Cardoso, G. Curio, D. L\"ust, T. Mohaupt and S.-J. Rey, 
``BPS spectra and nonperturbative gravitational couplings in 
$N=2$, $N=4$ supersymmetric string theories",
Nucl. Phys. {\bf B464} (1996) 18, {\tt hep-th/9512129}.

\bibitem{kiritsis}
E. Kiritsis and  C. Kounnas, ``Perturbative and nonperturbative 
partial supersymmetry breaking: $N=4\to N=2\to N=1$",
Nucl. Phys. {\bf B513} (1997) 117,
{\tt hep-th/9703059}.


\bibitem{FGPW1} 
D.~Z.~Freedman, S.~S.~Gubser, K.~Pilch and N.~P.~Warner,
``Renormalization group flows from 
holography, supersymmetry and a  c-theorem,''
{\tt hep-th/9904017}.

\bibitem{WFGK} 
O.~DeWolfe, D.~Z.~Freedman, S.~S.~Gubser and A.~Karch,
``Modeling the fifth dimension with scalars and gravity,''
{\tt hep-th/9909134}.

\bibitem{kallosh}
R.~Kallosh and J.~Kumar,
``Supersymmetry enhancement of D-p-branes and M-branes,''
Phys. Rev.  {\bf D56} (1997) 4934, {\tt hep-th/9704189}.

\bibitem{cvso}
M.~Cvetic, S.~Griffies and S.~Rey,
Nucl. Phys.  {\bf B381} (1992) 301, {\tt hep-th/9201007}.
M.~Cvetic and H.~H.~Soleng,
Phys. Rept.  {\bf 282} (1997) 159, {\tt hep-th/9604090}.


\bibitem{bs} I. Bakas and K. Sfetsos, ``States and curves of 
five-dimensional gauged supergravity", Nucl. Phys. {\bf B573}
(2000) 768, {\tt hep-th/9909041}.

\bibitem{bbs}
I. Bakas, A. Brandhuber and K. Sfetsos, ``Domain walls 
of gauged supergravity, M-branes, and algebraic curves", {\tt
hep-th/9912132}
(Adv. Theor. Math. Phys. in press). 


\bibitem{ioa} I. Bakas and Q-Han Park, ``Gravitational dressing of 
massive soliton theories", Phys. Lett. {\bf B387} (1996) 707,
{\tt hep-th/9607243}.

\bibitem{wald}
R. Wald, `` Dynamics in Nonglobally Hyperbolic, Static Space-Times,'',
J. Math. Phys. {\bf 21} (1980) 2802.

\bibitem{homa}
G.T.Horowitz and D. Marolf,
``Quantum probes of space-time singularities'',
Phys. Rev. {\bf D52} (1995) 5670,
{\tt gr-qc/9504028}.

\bibitem{ishibashi}
A.~Ishibashi and A.~Hosoya,
``Who's afraid of naked singularities? 
Probing timelike singularities  with finite energy waves,''
Phys.\ Rev.\  {\bf D60} (1999) 104028, 
{\tt gr-qc/9907009}.


\bibitem{BrS2}
A.~Brandhuber and K.~Sfetsos,
``Non-standard compactifications with mass gaps and Newton's law,''
JHEP {\bf 9910} (1999) 013, {\tt hep-th/9908116}.

\bibitem{DeWotsei}
O.~DeWolfe and D.~Z.~Freedman,
``Notes on fluctuations 
and correlation functions in holographic  renormalization group flows,''
{\tt hep-th/0002226}.

\bibitem{aft}
G.~Arutyunov, S.~Frolov and S.~Theisen,
``A note on gravity-scalar fluctuations in holographic RG flow 
geometries,''
{\tt hep-th/0003116}.


\bibitem{Cohen}
A.G. Cohen and D.B. Kaplan,
``Solving the hierarchy problem with noncompact extra dimensions,''
Phys. Lett.  {\bf B470} (1999) 52, 
{\tt hep-th/9910132}.

\bibitem{BrSf}
A. Brandhuber and K. Sfetsos,
``An N = 2 gauge theory and its supergravity dual'', {\tt
hep-th/0004148}.

\bibitem{tan} N. Deo, S. Jain and C.-I Tan, ``String distributions above
the Hagedorn energy density", Phys. Rev. {\bf D40} (1989) 2626.

\bibitem{djnt}
N. Deo, S. Jain, O. Narayan and C.-I Tan, ``Effect of topology on the 
thermodynamic limit for a string gas", Phys. Rev. {\bf D45} 
(1992) 3641. 
 

\bibitem{dvali} G. Dvali and M. Shifman, ``Surviving on the slope: 
supersymmetric vacuum in theories where it is not supposed to be", 
Phys. Lett. {\bf B454} (1999) 277, {\tt hep-th/9901111}.

\bibitem{gross} D. Gross, M. Perry and L. Yaffe, ``Instabilities of
flat space at finite temperature", Phys. Rev. {\bf D25} (1982) 330.

\bibitem{book} H. Davis, ``Introduction to nonlinear differential and
integral equations", Dover publications, New York, 1962.

\bibitem{abkr} 
S.A. Abel, J.L.F. Barbon, I.I. Kogan and E. Rabinovici,
``Some thermodynamical aspects of string theory", {\tt hep-th/9911004},
and ``String thermodynamics in D-brane backgrounds", JHEP {\bf 9904}
(1999)
015, {\tt hep-th/9902058}.

\bibitem{afian}
S.A. Abel, K. Freese and I.I. Kogan, ``Hagedorn inflation of D-branes",
{\tt hep-th/0005028}.


\end{thebibliography}
\end{document}